\documentclass[12pt]{article}

\usepackage{latexsym}

\begin{document}


\vskip 1truecm

\title{Neutron star in presence of torsion-dilaton field}

\author{Boyadjiev T.${}^{*}$, Fiziev P.${}^{**}$, Yazadjiev S.${}^{***}$\\
\\
{\footnotesize ${}^{*}$ Department of Analytical Mechanics,
Faculty of Mathematics and Computer Science,}\\
{\footnotesize Sofia University,}\\
{\footnotesize 5 James Bourchier Boulevard, Sofia~1164, }\\
{\footnotesize Bulgaria }\\
{\footnotesize E-mail: todorlb@fmi.uni-sofia.bg }\\
\\
{\footnotesize ${}^{**}$ Department of Theoretical Physics,
Faculty of Physics,}\\
{\footnotesize Sofia University,}\\
{\footnotesize 5 James Bourchier Boulevard, Sofia~1164, }\\
{\footnotesize Bulgaria }\\
{\footnotesize E-mail: fiziev@phys.uni-sofia.bg}\\
\\
{\footnotesize ${}^{***}$ Department of Theoretical Physics,
Faculty of Physics,}\\
{\footnotesize Sofia University,}\\
{\footnotesize 5 James Bourchier Boulevard, Sofia~1164, }\\
{\footnotesize Bulgaria }\\
{\footnotesize  E-mail: yazad@phys.uni-sofia.bg}
}

\maketitle

\begin{abstract}

We develop the general theory of stars in Saa's model of gravity with
propagating torsion and study the basic stationary state of neutron star.
Our numerical results show that the torsion force decreases the role of
the gravity in the star configuration leading to significant changes in the
neutron star masses depending on the equation  of state of star matter.
The inconsistency of the Saa's model with Roll-Krotkov-Dicke and
Braginsky-Panov experiments is discussed.

\noindent{PACS number(s): 04.40.Dg,04.40.-b,04.50.+h}
\end{abstract}

\sloppy
\renewcommand{\baselinestretch}{1.3} %
\newcommand{\sla}[1]{{\hspace{1pt}/\!\!\!\hspace{-.5pt}#1\,\,\,}\!\!}
\newcommand{\db}{\,\,{\bar {}\!\!d}\!\,\hspace{0.5pt}}
\newcommand{\partb}{\,\,{\bar {}\!\!\!\partial}\!\,\hspace{0.5pt}}
\newcommand{\dsla}{\partb}
\newcommand{\eql}{e _{q \leftarrow x}}
\newcommand{\eqr}{e _{q \rightarrow x}}
\newcommand{\ite}{\int^{t}_{t_1}}
\newcommand{\itz}{\int^{t_2}_{t_1}}
\newcommand{\itd}{\int^{t_2}_{t}}
\newcommand{\lfrac}[2]{{#1}/{#2}}
\newcommand{\sfrac}[2]{{ \,\,\hbox{${\frac {#1} {#2}}$}}}
\newcommand{\dV}{d^4V\!\!ol}
\newcommand{\ben}{\begin{eqnarray}}
\newcommand{\een}{\end{eqnarray}}
\newcommand{\la}{\label}


\section{Introduction}
In recent years the interest in scalar-tensor theories of gravity has been
renewed. One reason for this is the important role which these theories
play in the understanding of inflantionary epoch. On the other hand the
scalar-tensor gravitation (the so called "dilaton gravity") arises naturally
from the low-energy limit of the super-string theory \cite{Witten}, \cite{DT}.

The predictions of scalar-tensor theories may differ drastically from
these of general relativity. For example such a phenomenon --
"spontaneous scalarization" was recently discovered by Damour and
Esposito-Farese as a non-perturbative strong field effect
in a massive neutron star \cite{Damour}. Other interesting phenomenon is
"gravitational memory" of black holes proposed in \cite{Barrow}. The
"gravitational memory" in the case of boson stars was investigated in
\cite{TLS} (see also \cite{CS}).Their stability through cosmic history
using catastrophe theory was investigated in \cite{TSL}.

Many theories of gravity with propagating torsion involving a scalar field
have been proposed in the last decades, too \cite{HRRS}, \cite{HRR}, \cite{SG}.
In such theories contrary to the
usual Einstein-Cartan gravity \cite{Hehl1}-\cite{Hehl3},
there are long-range torsion mediated
interactions. Carrol and Field \cite{Carrol} have examined some observational
consequences of propagating torsion in a wide class of models involving
a scalar field. They conclude that for reasonable models the torsion could
be detected experimentally.

Recently a new interesting model with propagating torsion was proposed by Saa
\cite{Saa1}-\cite{Saa5}.
This model involves a non-minimally coupled scalar field as a
potential of the torsion of space-time. As one can see Saa's model
is very close to the dilaton gravity.

In the present article we investigate both analytically and numerically
a neutron star in the Saa's model and compare obtained results with these
in the general relativity. We also discuss new predictions of the theory
under consideration.

The paper is organized as follows.
In section 2 we consider briefly Saa's model.
In section 3 we give the necessary information for the vacuum solutions of
the field equations.
The equations determining static equilibrium solutions for a
neutron star are discussed in section 4.
Numerical results for the neutron star are discussed in section 5.
The stability of the neutron star is discussed via catastrophe theory in
section 6.
The inconsistency of the Saa's model with Roll-Krotkov-Dicke and
Braginsky-Panov experiments is discussed in section 7.

\section{The model with torsion-dilaton field}

Here we give a brief description of Saa's model. For more details one can see
\cite{Saa1}-\cite{Saa3}, \cite{Fiziev1}.

Consider four-dimensional Einstein-Cartan manifold  ${\cal M}^{(1,3)}$,
i.e. four-dimensional manifold equipped with metric \, $g_{\alpha\beta}$
and affine connection \, $\Gamma_{\alpha\beta}{}^\gamma$ \, with torsion
tensor \, ${\cal S}_{\alpha\beta}{}^\gamma$.

The main idea of articles \cite{Saa1}--\cite{Saa3} is to make the volume
form \,$d^4Vol $ \,compatible with the affine connection on the
Einstein-Cartan manifold \, ${\cal M}^{(1,3)} $ \,
via the compatibility condition:
\ben
\pounds_v\left(d^4V\!ol\right) = (\nabla_\mu v^\mu) d^4V\!ol,
\la{CC}
\een
where $\pounds_v$ is the Lie derivative along an arbitrary vector field $v$
and $\nabla_\mu$ is the covariant derivative with respect to the affine
connection. It turns out that compatibility condition (\ref{CC})
is fulfilled if and only if the torsion vector
$${\cal S}_{\alpha} = {\sfrac 2 {3}}  {{\cal S}_{\alpha\mu}}^
\mu$$  is potential, i.e. if there exists a potential $\Theta $, such that
\ben
S_{\alpha} = \nabla_\alpha \Theta \equiv \partial_\alpha \Theta.
\la{GrC}
\een
In this case Saa's condition (\ref{CC}) implies the form
\ben
d^4V\!ol = f(x) d^4 x =e^{-3\Theta}\sqrt{|g|} d^4 x
\la{Vol}
\een
of the volume element in Einstein-Cartan manifold. As it was pointed out
in \cite{Fiziev1} compatibility condition (\ref{CC}) leads to covariantly
constant scalar density $ f = e^{-3\theta}\sqrt{|g|}$ with respect to
the transposed connection
${(\Gamma^T)_{\alpha\beta}}^\gamma = \Gamma_{\beta\alpha}{}^\gamma $,
not with respect to the usual connection $\Gamma_{\alpha\beta}{}^\gamma$.
Therefore the Einstein-Cartan manifold for which compatibility condition
(\ref{CC}) is fulfilled was called transposed-equi-affine
and the  corresponding theory of gravity  --
transposed-equi-affine theory of gravity.

The most important mathematical consequence of the condition (\ref{CC})
which leads to new equations of gravity is the generalized Gauss' formula:
\ben
\int_{\cal M}d^4V\!ol \left(\nabla_\mu v^\mu\right) =
\int_{\partial{\cal M}} d^{3}\Sigma_\mu v^\mu.
\la{Gauss}
\een

The natural choice of the lagrangian density for gravity is:
\ben
{\cal L}_G = - {\sfrac {c} {2\kappa}}R =
- {\sfrac {c} {2\kappa}} \left(
\stackrel{\{\}}{R} + 6\nabla_\mu S^\mu + 12 S_\mu S^\mu
- {\tilde K}_{\mu\nu\lambda} {\tilde K}^{\mu\nu\lambda}\right),
\la{LG}
\een
$c$ being the velocity of light, $\kappa= 8\pi c^{-2} G$
being the Einstein constant, $G$ being the Newton constant.
Here $R = g^{\alpha\beta}R_{\alpha\beta}$ is the scalar curvature with
respect to the affine connection,
${\tilde K}_{\mu\nu\lambda}= K_{\mu\nu\lambda} + 2 g_{\mu[\nu} S_{\lambda]}$
is the traceless part of the contorsion:
${\tilde K}^\mu{}_{\mu\nu}={\tilde K}^\mu{}_{\nu\mu}\equiv 0$,
and  $\stackrel{\{\}}{R} $ is the scalar curvature with respect to the
Levi-Civita connection.

The traceless part of the torsion doesn't vanish only if
spin-non-zero matter presents.
In the present article we consider only spinless matter
(as we know from Einstein-Cartan theory of gravity,
the effects due to the spin become essential at density
over $10^{57} g/cm^3$ \cite{Hehl1} which is too far from the physics
in the stars).
Therefore we put ${\tilde K}_{\alpha
\beta\gamma}\equiv 0 $ and obtain a semi-symmetric affine connection:
\ben
{S_{\alpha\beta}}^\gamma = S_{[\alpha}\delta_{\beta]}^\gamma.
\la{SAT}
\een
In this case we have:
\ben
{\cal L}_G = - {\sfrac {c} {2\kappa}}R =
- {\sfrac c {2\kappa}} \left(
\stackrel{\{\}}{R} + 6\nabla_\mu S^\mu + 12 S_\mu S^\mu
\right).
\la{LG1}
\een

Denoting the lagrangian density for the matter by ${\cal L}_M $ and using
the volume element (\ref{Vol}) we write down the action of gravity and
matter in the form:
\ben
{\cal A} ={\cal A}_G + {\cal A}_M = -{\sfrac c {2\kappa}}\int\!\dV \,R
 + {\sfrac 1 {2c}}\int\!\dV {{\cal L}_M}.
\la{Action}
\een
Due to the new Gauss' formula (\ref{Gauss}) the term $ 6 \nabla_{\mu}S^{\mu}$
in the lagrangian (\ref{LG1}) gives a surface term in the action integral
(\ref{Action}) and doesn't contribute to the equations of motion.
Hence, these equations may be derived from the modified action:
\ben
\tilde{\cal A}= -{\sfrac c {2\kappa}}\int\!\dV\,\left(\stackrel{\{\}}{R}
 + 12{S^\mu}S_{\mu}\right) + {\sfrac 1 {2c}}\int\!\dV\,
{\cal L}_M .
\la{TildeA}
\een

This action is very close to the one of the dilatonic gravity arising from
low-energy limit of the superstring theory.
Two essential differences between our case and the dilatonic one are:
1) the matter action includes the dilaton-like term $e^{-3\theta}$ which
arises in a natural way, as a part of the volume element of space-time, and
2) the sign before the term $12S^{\mu}S_{\mu}$.
Following the above described reasons we call the field $\Theta$,
which originates from the space-time torsion and plays
the role of the dilaton
field in Saa's model, ''a torsion-dilaton field".

Taking variations with respect to the metric $g_{\alpha\beta}$ and 
torsion-dilaton field $\Theta$, and using the generalized Gauss' formula,
we obtain the following equations of motion for the geometrical fields
$g$ and $\theta$:
\ben
G_{\mu\nu} +\nabla_\mu \nabla_\nu \Theta - g_{\mu\nu} \Box \Theta =
{\sfrac \kappa {c^2}}T_{\mu\nu},\nonumber\\
\Box\Theta  = {\sfrac \kappa {c^2}}
\left({\cal L}_M -{\sfrac 1 3}{\delta{\cal L}_M \over \delta \Theta}\right)
-{\sfrac 1 2} R.
\la{GFE}
\een
Here $G_{\mu\nu} = R_{\mu\nu}-{\sfrac 1 2} g_{\mu\nu}$ is the Einstein tensor
for the affine connection, its trace is $G=g^{\mu\nu}G_{\mu\nu}=-R$;
$T_{\mu\nu} = {\lfrac {\delta{\cal L}_M} {\delta g^{\mu\nu}}}$
is the symmetric energy-momentum tensor of the matter ; its trace is
$T=g^{\mu\nu}T_{\mu\nu}$ and  $\nabla_\sigma S^\sigma
 = g^{\mu\nu}\nabla_\mu \nabla_\nu \Theta= \Box\Theta$.
From the first equation of the system (\ref{GFE}) it follows that:
\ben
R = -3\Box\Theta  - {\kappa\over c^2}T
\la{SCurv}
\een
Then combining this result with the second equation of the system (\ref{GFE})
we obtain:
\ben
\nabla_\sigma S^\sigma= \Box\Theta= -{\sfrac {2\kappa} {c^2}}
\left({\cal L}_M -{\sfrac 1 3}{\delta{\cal L}_M \over \delta \Theta}
+{\sfrac 1 2}T\right).
\la{DivS}
\een
The equation (\ref{DivS}) shows that under proper boundary conditions, and
in the presence only of spinless matter, the  torsion-dilaton field
$\Theta$ is completely determined by the matter distribution.
Further on, as a basic system we will use the system:
\ben
G_{\mu\nu} +\nabla_\mu \nabla_\nu \Theta - g_{\mu\nu}\Box\Theta =
{\sfrac \kappa {c^2}}T_{\mu\nu},\nonumber\\
\nabla_\sigma S^\sigma= \Box\Theta= -{\sfrac {2\kappa} {c^2}}
\left({\cal L}_M -{\sfrac 1 3}{\delta{\cal L}_M \over \delta \Theta}
+{\sfrac 1 2}T\right).
\la{SYS}
\een
From this system one can derive (using Bianchi identity) the
differential consequence:
\ben
\nabla_\sigma T_\alpha^\sigma + T^\sigma_\alpha S_\sigma =
{\sfrac {c^2} {2\kappa}}R S_\alpha
\la{DivT}
\een
which is a generalization of the well-known conservation law
${\stackrel{ \{\} }
{\nabla} }{}_\sigma T_\alpha^\sigma=0$  in general relativity.

To have a complete set of dynamical equations one has to add to the above
relations the equations of motion of the very matter. For the purpose of the
present article we need to consider only a perfect fluid. Its theory was
recently described in \cite{Fiziev1}. Here we give the basic results.

The continuity condition describing the conservation of the fluid matter
can be written in the form:
\ben
\int_{\partial \Delta^{(1,3)}} d^3\Sigma_\alpha \,n(x) u^\alpha(x) = 0,
\la{IntCE}
\een
where $u^\alpha(x)$ is the fluid four-velocity, normalized by the relation
$g_{\alpha\beta}u^{\alpha}u^{\beta} = 1$, ${n}(x) $ is properly defined
a fluid density, $d^3\Sigma_\alpha $ is a proper three dimensional
surface element depending on the choice of the volume element via the Gauss'
formula, and  $\Delta^{(1,3)} $ is an arbitrary domain.

Considering the volume element (\ref{Vol}) as an universal one we must
use it in
the continuity condition, too. Therefore according to the generalized
Gauss' formula we can rewrite relation (\ref{IntCE}) in the form of
a continuity equation of autoparallel type:
\ben
\nabla_\alpha \left( n(x) u^\alpha(x) \right) = 0.
\la{ACE}
\een

We take the lagrangian of the fluid with internal pressure $p$ in the usual
form:
\ben
{\cal L}_\mu = - \varepsilon = -n c^2 - n\Pi,
\la{FLagrangian}
\een
where  $\Pi $ is the elastic potential energy of the fluid ; $\db \Pi =
- p d({\frac 1 n})$ and the symbol " $\db$ " means that the corresponding
differential form isn't exact.
Taking into account the relation
${\cal L}_\mu -{\sfrac 1 3}{\delta{\cal L}_\mu \over \delta \Theta} = p$
it's not difficult to obtain the equations of motion for geometrical fields
$g_{\alpha\beta}$  and $\Theta$ in presence of perfect fluid:
\ben
G_{\mu\nu} + (\nabla_\mu \nabla_\nu - g_{\mu\nu}\Box)\Theta &=&
{\sfrac \kappa {c^2}}\biggl((\varepsilon + p)u^\mu u^\nu
- p\,g^{\mu\nu}\biggr),\nonumber\\
\Box\Theta  &=& - {\sfrac \kappa {c^2}} (\varepsilon - p);
\la{GFE_Flu}
\een
In addition one can show that:
\ben
\nabla_\sigma T_\alpha^\sigma &=&
(\varepsilon + p)\left(\delta^\sigma_\alpha-u^\sigma u_\alpha\right)S_\sigma,
\,\,\,\,\,\,\,\,\,\,\,\,\hbox{or}\nonumber\\
{\stackrel{ \{\} } {\nabla} }{}_\sigma T_\alpha^\sigma &=&
3(\varepsilon + p)\,u^\sigma u_\alpha S_\sigma.
\la{DivT_Flu}
\een
Making use of (\ref{DivT_Flu}) and of the continuity condition  (\ref{ACE})
one can
obtain the equations of motion of the perfect fluid (just as in  general
relativity):
\ben
(\varepsilon + p) u^\beta {\stackrel{{\{\}}}{\nabla}}_\beta u_\alpha =
\left(\delta^\beta_\alpha - u_\alpha u^\beta \right)
{\stackrel{{\{\}}}{\nabla}}_\beta p.
\la{GFluEM}
\een
The equations  (\ref{DivT_Flu})  are equations of a geodesic type.
In particular, considering dust  matter ($p\equiv 0$)  we have:
\ben
 u^\beta {\stackrel{{\{\}}}{\nabla}}_\beta u_\alpha = 0,
\la{EM_dust}
\een
i.e. we can conclude (just as in general relativity)  that a test particle
in the theory under consideration will move on a geodesic line.
We will need this conclusion in the next sections.
For more details concerning the relativistic perfect fluid in the theory
under consideration we refer to \cite{Fiziev1}.

\section{Spherically symmetric vacuum solution}
The asymptotic  flat, static and spherically symmetric general solutions of
the vacuum  geometrical field equations  (\ref{GFE}) are known \cite{Brans},
\cite{Zannias}.
In Schwarzschild's coordinates they are described as a two parameter --
$\{K, a\}$ family of solutions\footnote{We use asymptotic conditions
$\nu \rightarrow 0,\Theta \rightarrow  0$ at $r \rightarrow  \infty$
without loss of generality.}:

\ben
g_{00} = e^{\nu}, \nonumber \\
\nonumber \\
r ={\sfrac 1 2} a e^{(3K - 1)\nu\over 2 }
\sinh ^{-1}\left(\sfrac {\rho\nu} 2\right), \nonumber \\
\nonumber \\
g_{11} = e^{\lambda} = \left({\sfrac {1 + \delta} 2} e^{-\rho\nu\over 2 }
+ {\sfrac {1 - \delta} 2} e^{\rho\nu\over 2}\right)^{-2},  \nonumber  \\
\nonumber \\
\rho = \sqrt{ 3\left(K - {\sfrac 1 2}\right)^2 + {\sfrac 1 4}}, \nonumber  \\
\nonumber \\
\delta ={\sfrac {3K - 1} \rho }  \nonumber  \\
\nonumber \\
\xi = {\sfrac 1 2} \nu\,'  =  {\sfrac 1 2}\, {\sfrac a \rho}\, {\sfrac
1 {r^2}}\, e^{(3K - 1)\nu \over 2 }\,  e^{\lambda \over 2} \nonumber   \\
\nonumber \\
S_{r} =\Theta'= K \xi.
\la{VS}
\een
Here and further on the prime denotes a differentiation with respect to the
variable $r$.
All quantities in formulae (\ref{VS}) are represented as functions of the
variable $\nu$. This is the most convenient form of the vacuum solutions.

The parameter $K$ presents the ratio of the torsion force
(as defined in \cite{Fiziev1}) and the gravitational one:
$K = { {S_r} / ({\sfrac 1 2} \nu\,') }$.
In the case when $K = 0$ we have the usual torsionless Schwarzschild's
solution
and $a \equiv r_g$ is the standard gravitational radius $r_g$.

In the model under consideration the value of the fundamental parameter
of the theory $K$ (which is constant in vacuum)
is not an independent integration constant.
Instead, we shall show that it is determined by the total mass of the star,
or by its radius
and depends on the matter distribution,
on the equation of state of the star's matter, and so on
via the solution of the full system of equation of the star's state.

The parameter $a$ is positive ($a > 0$), and  may take arbitrary values.
It is related to the total mass of the star, too.

The asymptotic behaviour of the solution is:
\ben
g_{00} \sim 1 - {2M_{{}_{Kepler}}G\over c^2r},   \\ \nonumber
\\
g_{11}  \sim 1 +  {2G(M_{{}_{Kepler}} - M_{\theta}) \over c^2r},
\la{AB}
\een
where $M_{{}_{Kepler}} = {1\over 2} {c^2\over G} {a\over \rho}$
describes the asymptotic dependence of $g_{00}$ on the variable $r$,
and the mass $M_{\theta} = 3{c^2 \over G} \lim_{r\to\infty} r^2 S_{r}$
describes the asymptotic dependence of the torsion-dilaton potential
$\Theta$ on $r$ (or the asymptotic of $S_{r}$).
The mass $M_{\theta}$  may be considered as a source
of torsion-dilaton force and is analogous to the
"scalar mass'' introduced in \cite{Lee}.
As one can see the scalar mass $M_{\theta}$ depends on $K$.

In the model under consideration, a test particle moves along geodesic lines.
Therefore, the keplerian mass measured by a test particle in the
asymptotic region of space-time will be the mass
$ M_{{}_{Kepler}} = {c^2 \over G}\lim_{r \to\infty}
{1\over 2} r^2 \nu\,'e^{\nu}$.
As we will see in the next section, the mass $M_{{}_{Kepler}}$ is positively
defined.
The appearance of two masses in the asymptotic solution is related to the
violation of the strong equivalence principle
(the weak equivalence principle for test particles is not violated).
In our case the ratio  $M_{\theta}$
to $M_{{}_{Kepler}}$ is just $3K$ and depends on the ratio $K$ of the
torsion-dilaton force to gravitational one in vacuum.
As in Brans-Dicke theory one can define so called tensor mass $M_T$,
which is the mass measured  by a test particle in Einstein frame, i.e. a test
particle moving along  a geodesic in space-time with metric
$\tilde g_{\mu\nu} = e^{-3\Theta} g_{\mu\nu}$. It's not difficult to see that
$M_T$ and $M_{{}_{Kepler}}$  are related by the formula:
\ben
M_T =\left(1 - {3\over 2}K\right)M_{{}_{Kepler}}.
\la{TMass}
\een
In the next section we show that the tensor mass $M_T$
is also positively  defined. The existence of two positively defined masses
$M_{Kepler}$ and $M_{T}$ makes the question  about the energy of
the geometric-field complex ${g,\Theta}$ subtle - which  of them is
the true measure of the energy of the star?

It turns out that one has to choose the tensor mass $M_T$. The reasons for
this we will comment shortly in the next section. For detailed consideration
of the question of the star mass definition in presence of a scalar field
we refer the reader to \cite{Whinnett}.

\section{The Basic equations for a star}
\subsection{General considerations}
Here we will discuss some general  properties of the system of equations
of the star without specifying the matter's equation of state.

The system of the equation (\ref{GFE_Flu} ) for geometrical fields
$g$  and $ \Theta$ can be rewritten in the form:
\ben
{\stackrel{\{\}}{G}}{}^\nu_\mu + 3{\Sigma}^{\nu}_{\mu} =
{\sfrac \kappa {c^2}} T^{\nu}_{\mu}, \nonumber \\
{\stackrel{\{\}}{\nabla}}_{\sigma} S^{\sigma} - 3 S_{\sigma}S^{\sigma} =
 - {\sfrac \kappa {c^2}} (\epsilon - p),  \nonumber \\
T^{\nu}_{\mu} = (\epsilon + p)u^{\nu}u_{\mu} - p g^{\nu}_{\mu},
\la{E1}
\een
where $\stackrel{\{\}}{\nabla}$ is the covariant derivative
with respect to the Levi-Civita connection,
${\stackrel{\{\}}{G}}{}^\nu_\mu$ is the corresponding Einstein's tensor and
\ben
{\Sigma}^{\nu}_{\mu} = { \stackrel{\{\}}{\nabla}}_{\mu}S^{\nu} +
S_{\mu}S^{\nu}
- g^{\nu}_{\mu}{ \stackrel{\{\}}{\nabla}}_{\sigma}S^{\sigma} +
g^{\nu}_{\mu} S_{\sigma}S^{\sigma}.
\een

In this paper we restrict ourselves with consideration of the static
and spherically symmetric case. Hence, the metric has the form
\ben
ds^2 = e^{{\nu}(r)}(c\,dt)^2  - e^{{\lambda}(r)}dr^2  - r^2 (d{\theta}^2 +
\sin^2(\theta)d{\varphi}^2 )
\een
in  which the functions  $\nu = {\nu}(r)$, $\lambda = {\lambda}(r) $
depend only on the Schwarzschild's radial coordinate  $r$  and the
torsion-dilaton field  $\Theta$  depends only on  $r$, too.

In this case we obtain the following equations for the functions
$\nu, \lambda, \Theta$, and $p$:
\ben
- e^{- \lambda}
\left({\sfrac 1 {r^2}}-{\sfrac {\lambda'} r}\right)+{\sfrac 1 {r^2}}=
{\sfrac \kappa {c^2}} \varepsilon - 3{\Sigma}^{0}_{0},  \nonumber  \\
- e^{- \lambda}\left({\sfrac {\nu\,'} {r}}+{\sfrac 1 {r^2}}\right)+
{\sfrac 1 {r^2}}=
- {\sfrac \kappa {c^2}}p - 3{\Sigma}^{1}_{1},  \nonumber   \\
- {\sfrac 1 2} e^{-\lambda} \left({\nu\,''}+{\sfrac 1 2}{{\nu\,'}}^2 +
{\sfrac {\nu\,' - \lambda'}  r}  - {\sfrac {\nu\,' \lambda'} 2}\right)
= - {\sfrac \kappa {c^2}}p  - 3{\Sigma}^{2}_{2}, \nonumber    \\
e^{-\lambda} \left( {S_{r}'} + \left({\sfrac {\nu\,' - \lambda'} 2} +
{\sfrac 2 r} \right) S_{r} - 3{S_{r}}^2 \right) =
{\sfrac \kappa {c^2}} (\varepsilon - p),
\nonumber \\
{p'} = - {\sfrac 1 2} (\varepsilon + p) {\nu\,'}, \nonumber \\
p = p(\varepsilon).
\la{SSE}
\een
Here $p = p(\varepsilon)$ is the matter's equation of state
and correspondingly:
\ben
{\Sigma}^{0}_{0} = \left({S_{r}'} - {\sfrac 1 2}{\lambda'} S_{r} +
{\sfrac 2 r}S_{r}  -  {S_{r}}^2 \right) e^{-\lambda},  \nonumber \\
{\Sigma}^{1}_{1} = \left({\sfrac 1 2}{\nu\,'} S_{r}  +  {\sfrac 2 r}S_{r}
 -  2{S_{r}}^2 \right) e^{-\lambda}, \nonumber \\
{\Sigma}^{2}_{2}  = \left({S_{r}'} + {\sfrac 1 2}{\nu\,'} S_{r} -
{\sfrac 1 2}{\lambda'}S_{r} +  {\sfrac 1 r}S_{r} - {S_{r}}^2 \right)
e^{-\lambda}.
\een
The second equation of the system (\ref{SSE})
(i.e. $ \stackrel{\{\}}{G^{1}}_{1} + 3{\Sigma}^{1}_{1} =
{\kappa \over c^2} T^{1}_{1} $  ) may be considered as
a constraint creating a relation between $S_{r}, {\nu\,'}$  and
$e^{\lambda} $, namely:
\ben
e^{\lambda} = {{1 + r{\nu\,'} - 6rS_{r}  -  {3 \over 2}r^2{\nu\,'}
S_{r}  +  6r^2{S_{r}}^2} \over {1 + {\sfrac \kappa {c^2}}pr^2 }}.
\een
Using this relation we can put our system (\ref{SSE}) in a normal form:
\ben
{\nu\,'} = 2 \xi, \nonumber \\
{\Theta'} = S_{r}, \nonumber \\
\nonumber \\
{\xi'} = -{\sfrac \xi r} + \left ({\sfrac {2\kappa} {c^2}}\varepsilon -
{\sfrac \xi r}  - {\sfrac\kappa {c^2}}(\varepsilon  - p){r \xi} \right)
 e^{\lambda}, \nonumber \\
{S_{r}'} = -{\sfrac {S_{r}} r} + \left({\sfrac \kappa {c^2}}(\varepsilon
- p)  - {\sfrac {S_{r}} r}  -  {\sfrac \kappa {c^2}}(\varepsilon  - p)
{r S_{r}} \right) e^{\lambda},  \nonumber \\
p' = - (\varepsilon  +  p)\xi, \nonumber \\
p = p(\varepsilon),  \nonumber \\
\nonumber \\
e^{\lambda} = {{1 + 2r{\xi} - 6rS_{r}  -  3r^2{\xi}
S_{r}  +  6r^2{S_{r}}^2} \over {1 + {\sfrac \kappa {c^2}}pr^2 }}.
\la{NF}
\een
It's seen that the first two equations are separated, and the rest generate
a subsystem independent of them.

The equations (\ref{NF})  must be solved with proper initial and boundary
conditions. From a physical point of view the solutions regular at the center
are the most interesting. The regularity means that there exists
a local lorentzian system in neighborhood of the center, i.e. $e^{\lambda(0)}
= 1 $ and the pressure is finite at $r=0$. Hence, we have $lim_{r\to 0}\,
\,r\xi(r) = 0$, otherwise, as it may be seen from equation for $p$, the
pressure will have at least logarithmic singularity at the center. On the
other hand  the condition  $e^{\lambda(0)} = 1 $ requires that $lim_{r\to 0}\,
\,rS_{r}(r) = 0$, too. The expansion of the equations for $\xi$ and $S_{r}$
around the center is:
\ben
{\xi'} = - {2\xi  \over r }, \nonumber \\
{S_{r}'} = -{2S_{r} \over r}.
\een
Hence, the behaviour of $\xi$  and $S_{r}$  around  $r = 0$ is ${constant
\over r^2}$. To fulfill the above restrictions at $r \to 0$ we must put
$constant = 0$. Hence, we obtain $\xi(0) = 0$, $S_{r}(0) = 0$.
As a final result these considerations imply the following initial conditions:
\ben
{\xi}(0) = 0 ,\,\,\,\,{S_{r}}(0) = 0, \,\,\,\,p(0) = p_{c}, \,\,\,\,
 ({\varepsilon}(0) = {\varepsilon}_{c}).
\een

At the star surface $r = R$ we have to match interior solution with
the exterior (vacuum) solution.
We will consider the model of the star without surface tension,
hence $p(R) = 0$.
Then the matter distribution must be continuous at the surface of the star
and one can show that $\xi$ and $S_{r}$ must be continuous at $r = R$.
Obviously $\nu$  and $\Theta$  must be continuous at
the star surface, too. Using matching conditions:
\ben
\xi(R) = {\xi}^{ext}(R),   \nonumber \\
S_{r}(R) = {S_{r}}^{ext}(R),
\een
we can obtain the vacuum solutions parameters  $K$ and $a$  as functions of
$\varepsilon_{c}$, i.e.
\ben
K = K(\varepsilon_{c}), \,\,\,\, a = a(\varepsilon_{c}).
\een

For arbitrary values $\nu_{c} = \nu{(0})$  and $\Theta_{c} = \Theta{(0)} $,
$\nu$  and $\Theta$  will not fulfill the matching conditions:
\ben
\nu(R) = {\nu}^{ext}(R), \nonumber \\
\Theta(R) = {\Theta}^{ext}(R).
\een
Therefore, the separated equations ${\nu\,'} = 2 \xi$ and ${\Theta'}= S_{r} $
must be solved under proper initial conditions in the following form:
\ben
\nu_{c} = {\nu}^{ext}(R) - \int_0^R 2 \xi(r)\, dr, \nonumber \\
\Theta_{c} = {\Theta}^{ext}(R) - \int_0^R \Theta(r)\, dr.
\een

As a result we obtain all parameters $K$, $a$, $\nu_{c}$, $\Theta_{c}$,
$R$  as functions only of the central density $\varepsilon_{c}$.
Hence, the whole geometry of the space-time in vacuum and in the star
is completely determined by the matter which carries
{\em only} the same properties described by mass, matter density,
pressure, equation of state and so on which are
familiar from the general relativity.
A very important feature of the model under consideration is that
we are not forced to assign to the matter new properties, charges,
or something else. Nevertheless we have a new geometric field
(the torsion-dilaton field $\Theta$) the physical problem is well defined
by the usual physical properties of the matter.

Let's go back to the subsystem:
\ben
{\xi'} = -{\sfrac \xi r} + \left({\sfrac {2\kappa} {c^2}}\varepsilon -
{\sfrac \xi r}  - {\sfrac \kappa {c^2}}(\varepsilon  - p) r \xi \right)
 e^{\lambda},
\nonumber \\
{S_{r}'} = -{\sfrac {S_{r}} r} + \left({\sfrac \kappa {c^2}}(\varepsilon
- p)  - {\sfrac {S_{r}} r}  -  {\sfrac \kappa {c^2}}(\varepsilon  - p)
r S_r \right) e^{\lambda},  \nonumber \\
p' = - (\varepsilon  +  p)\xi,  \nonumber \\
p = p(\varepsilon),  \nonumber \\
\nonumber \\
e^{\lambda} = {{1 + 2r{\xi} - 6rS_{r}  -  3r^2{\xi}
S_{r}  +  6r^2{S_{r}}^2} \over {1 + {\kappa \over {c^2}}pr^2 }}.
\la{SubS}
\een

We can't define a local gravitational mass in the form
$\,\, m_{GR}(r) = {c^2 \over 2G}r(1 - e^{-\lambda})\,\, $,
as in general relativity because in our case
$\,\,m_{GR}(r)\,\,$ is in general not positively defined
(see for example the vacuum solution).
In the theory under consideration we define the local mass as
$\,\, m_{\nu}(r) = {c^2 \over G} r^2\xi(r)\,\,$.
This local mass is connected with local Keplerian mass $m_{{}_{Kepler}}(r)$
(the mass measured by a non-self gravitating test particle in a circular
geodesic orbit with radius $r$ \cite{Whinnett}) by the relation
$$m_{{}_{Kepler}}(r)=m_{\nu}(r)e^{\nu}.$$
Similarly, we can define a local scalar mass
$\,\, m_{\theta}(r) = 3{c^2 \over G} r^2S_{r}(r) \,\, $.
We can introduce also a local tensor mass $m_T(r)$ \cite{Whinnett} as
$$m_T(r) ={\left(m_{\nu}(r) - {1\over 2}m_{\Theta}\right)e^{\nu}
\over \left(1 -{Gm_{\Theta}\over c^2 r}\right)}.$$
The initial conditions imply
$m_{\nu}(0) =m_{{}_{Kepler}}(0)=m_{\theta}(0) =m_T(0)= 0$.

Now the system (\ref{SubS}) can be rewritten in terms of the masses
$m_{\nu}(r)$ and $m_{\theta}(r)$:
\ben
m_{\nu}' = \left(1 - \left( 1 + {\sfrac \kappa {c^2}}(\varepsilon - p)r^2
\right)e^{\lambda} \right){\sfrac {m_{\nu}} r} + {\sfrac {16\pi} {c^2}}
r^2\varepsilon  e^{\lambda}, \nonumber \\
m'_{\theta} = \left(1 - \left( 1 + {\sfrac \kappa {c^2}}(\varepsilon - p)r^2
 \right)e^{\lambda} \right){\sfrac {m_{\theta}} r}+{\sfrac {24\pi} {c^2}}r^2
(\varepsilon  - p)e^{\lambda},   \nonumber \\
p' = -{\sfrac G {c^2}}(\varepsilon + p){\sfrac {m_{\nu}(r)} {r^2}} \nonumber \\
p = p(\varepsilon),  \nonumber \\
 \nonumber \\
e^{\lambda} ={{1 + {2G \over c^2r} (m_{\nu} - m_{\theta}) - {{G^2}
 \over c^4 r^2}
m_{\theta}(m_{\nu} - {2 \over 3}m_{\theta}) } \over {1 + {\kappa \over c^2}pr^2}}.
\la{OV}
\een
This system is a generalization of Tolman-Oppenheimer-Volkoff's one for
a star in general relativity \cite{HTWW}.
Using the first and the second equation,
it's not difficult to show that  $m_{\nu}(r)$ and $ m_{\theta}(r)$
are positively defined. Indeed, taking into account
regularity at the center we obtain:
\ben
m_{\nu} =  e^{A(r)}\int_0^r e^{-A(r)}  {\sfrac {16\pi} {c^2}}
\varepsilon r^2 dr,
\nonumber \\
m_{\theta} =  e^{A(r)}\int_0^r e^{-A(r)} {\sfrac {24\pi} {c^2} }(\varepsilon
- p) r^2 dr,
\een
where  $$ A(r) = \int_0^r  {\sfrac { 1 - \left( 1  + {\kappa \over c^2}
(\varepsilon - p)r^2  \right)e^{\lambda} }  r} dr.$$
In the same way from the above equations we have:
\ben
(m_{\nu} - m_{\theta})' =  \left(1 - \left( 1 + {\sfrac \kappa {c^2}}
(\varepsilon - p)r^2  \right)e^{\lambda} \right)
{\sfrac {m_{\nu}  - m_{\theta}} r}
 - {\sfrac {8\pi} {c^2}} r^2 (\varepsilon  - 3p)e^{\lambda}.
\een
Solving this equation with an initial condition $(m_{\nu} - m_{\theta})(0)
= 0$ we obtain :
\ben
m_{\nu} - m_{\theta} = -  e^{A(r)}\int_0^r e^{-A(r)} (\varepsilon - 3p)dr.
\een

Hence, it's seen that ${m_{\nu} - m_{\theta}} \leq 0$
inside and outside the matter if ${\varepsilon - 3p} \geq 0 $.
In other words we obtain  for $k(r)={1 \over 3} {m_{\theta}(r)
\over m_{\nu}(r)}$  that $k\geq {1 \over 3}$ and $k$  takes  a
value ${1 \over 3}$ when
the matter is ultrarelativistic ($\varepsilon = 3p$).
The parameter $k$ takes its maximum value $ {1 \over 2 }$
in the case of nonrelativistic matter ($\varepsilon \gg p$).
If we assume following Zel'dovich \cite{Zel'dovich1}, \cite{ZN}  that
$\varepsilon < 3p $ may happen, then in general
the vacuum value of k which is just $K = k(R)$ may change its
sign passing through the zero at $\varepsilon = p$. For realistic
equations of state we obtain $K\in[{1 \over 3},{1 \over 2}]$.
Therefore we have
$M_{T}\in [{\sfrac 1 4}{M_{{}_{Kepler}}},{\sfrac 1 2}{M_{{}_{Kepler}}}]$
(see (\ref{TMass})).

For completeness we will give an expression which is a generalization of the
well-known Tolman's formula \cite{Tolman}. From equations  (\ref{E1})
we have
\ben
{\stackrel{\{\}}{R}}{}^0_0 + 3\stackrel{\{\}}{\nabla}_{0}S^{0} =
{\sfrac \kappa {c^2}}
\left(T^{0}_{0}  - {\sfrac 1 2}T\right)  - {\sfrac 3 2}\Box\Theta.
\een
In the static and spherically symmetric case, one may show that the following
relation is fulfilled:
\ben
{\stackrel{\{\}}{R}}{}^0_0 + 3\stackrel{\{\}}{\nabla}_{0}S^{0} =
{1 \over e^{-3\Theta}\sqrt{|g|} } {\partial}_ {\alpha}
\left(\sqrt{|g|}e^{-3\Theta}
g^{0\beta} {\stackrel{\{\}}{\Gamma}}_{0\beta}{}^\alpha \right),
\een
where
${\stackrel{\{\}}{\Gamma}}_{\alpha\beta}{}^{\gamma}=
{\gamma \brace \alpha \beta}$
are Christoffel symbols. Hence, we obtain
\ben
\int_{\cal M}\left({\stackrel{\{\}}{R}}{}^0_0 +
3\stackrel{\{\}}{\nabla}_{0}S^{0}\right)
e^{-3\Theta}\sqrt{|g|} d^{3}x  =  \oint_{{\Sigma}_{\infty}} g^{0\beta}
{\stackrel{\{\}}{\Gamma}}_{0\beta}{}^\alpha
e^{-3\Theta} \sqrt{|g|} d\Sigma_{\alpha} = 4\pi{\sfrac {M_{{}_{Kepler}}G} {c^2}}.
\een
Therefore for Keplerian mass we can write down
\ben
M_{{}_{Kepler}} = {\sfrac {c^2} {4\pi G}}
\int_{\cal M}\left({\stackrel{\{\}}{R}}{}^0_0 +
3\stackrel{\{\}}{\nabla}_{0}S^{0}\right)
e^{-3\Theta}\sqrt{|g|} d^{3}x = \nonumber\\
{\sfrac 1 {c^2}}\int_{\cal M}(2T^{0}_{0} - T) e^{-3\Theta}\sqrt{|g|} d^{3}x -
{\sfrac {c^2} {4\pi G}}{\sfrac 3 2} \int_{\cal M}\Box\Theta
e^{-3\Theta}\sqrt{|g|} d^{3}x.
\een
Taking into account that
$$ \int_{\cal M}\Box\Theta e^{-3\Theta} \sqrt{|g|} d^{3}x =
- {\sfrac {4\pi} 3}{\sfrac {M_{\theta} G} {c^2}}$$  and
$2T^{0}_{0} - T = \varepsilon + 3p $  we obtain:
\ben
M_{{}_{Kepler}} =  {\sfrac 1 {c^2}}\int_{\cal M} (\varepsilon + 3p)
e^{-3\Theta}\sqrt{|g|}d^{3}x  + {\sfrac 1 2} M_{\theta}.
\een
On the other hand, taking into account that
$$M_{\theta} = 3KM_{{}_{Kepler}} = {\sfrac 6 {c^2}}
\int_{\cal M}(\varepsilon~-~p) e^{-3\Theta}\sqrt{|g|} d^{3}x $$
one can rewrite the above formula in the form
\ben
M_{{}_{Kepler}} = {\sfrac 1 {1 - {3 \over 2} K} }{\sfrac 1 {c^2}}
\int_{\cal M} (\varepsilon + 3p)
e^{-3\Theta}\sqrt{|g|} d^{3}x\nonumber \\
= {\sfrac 4 {c^2}}\int_{\cal M} \varepsilon
e^{-3\Theta}\sqrt{|g|} d^{3}x =
{\sfrac {16\pi} {c^2}} \int_0^R \varepsilon e^{{(\lambda + \nu) \over 2} -
3\Theta}r^2 dr.
\een
Comparing this formula with (\ref{TMass}) we immediately obtain the
following expression for the tensor mass
\ben
M_{T}={1\over c^2}\int_{\cal M} (\varepsilon + 3p)
e^{-3\Theta}\sqrt{|g|} d^{3}x ={4\pi \over c^2}
\int_0^R (\varepsilon + 3p) e^{{(\lambda + \nu) \over 2} -
3\Theta}r^2 dr.
\een
From the explicit expressions for the masses we have
\ben
0\leq M_{T}\leq M_{{}_{Kepler}} \leq M_{\Theta}
\een
when matter satisfies the condition $\varepsilon - 3p \geq 0$.\\
The number of the particles in the theory under consideration  is given by
\ben
N=4\pi \int_0^R n(r)e^{{\lambda \over 2} - 3\Theta} r^2 dr
\een
where $n(r)$ is the particle density. We note that we consider cold matter
and  we have $\varepsilon=\varepsilon(n)$ , $p=p(n)$.
Therefore for the rest mass $M_{R}$  we have
\ben
M_{R}=m_{N}N =4\pi m_{N}\int_0^R n(r)e^{{\lambda \over 2} - 3\Theta} r^2 dr
\een
where $m_{N}$ is the nucleon mass.

The binding energy of the star $E_{B}$ depends on the choice of the star mass.
As it is shown in \cite{Whinnett} the right choice must satisfies the
following  physically reasonable conditions in the spherically symmetric case:

1) to be non-negative;

2) to be identically  zero in Minkowski space-time;

3) to be non-decreasing function of $r$;

4) the maximum of the mass to coincide with the maximum of the rest mass(
the number of the particles).

As it is seen from numerical calculations both Keplerian and tensor mass
satisfy the first three conditions but only the tensor mass satisfies the
fourth. Therefore binding energy must be calculated with respect to the
tensor mass:
\ben
E_{B}=M_{T} - M_{R}.
\een

\subsection{Neutron star model}
 First we  consider non-interacting neutron gas at zero temperature
\cite{Shapiro}, \cite{LL}.
The energy density, the pressure and the particle number density in a proper
normalization are given by:
\ben
\varepsilon = {m_{N}^4c^5 \over {3{\pi}^2{h}^3}} g(\mu), \nonumber \\
 p =  {m_{N}^4c^5 \over {3{\pi}^2{h}^3}} f(\mu),    \nonumber   \\
 n={m_{N}^3c^3 \over 3{\pi}^2h^3}{\mu}^{3\over 2}
\een
where
\ben
g(\mu) = {\sfrac 1 8} \left( 8\mu\sqrt{\mu + {\mu}^2 } - \sqrt{\mu  + {\mu}^2 }
(2\mu - 3) - 3ln(\sqrt{\mu}  + \sqrt{1 + \mu}) \right) \nonumber, \\
f(\mu) = {\sfrac 1 8} \left(\sqrt{\mu  + {\mu}^2 }(2\mu - 3) + 3ln(\sqrt{\mu}
+ \sqrt{1 + \mu})\right),
\la{EOS1}
\een
$\mu = ({q_{Fermi} \over {m_{N}c }})^2$,
$q_{Fermi}$ being  the Fermi's momentum, $m_{N}$ being  the neutron mass.

We are interested in the difference between the predictions of the theory
under consideration and of the general relativity.
For this purpose the equation of
state for a non-interacting neutron gas is sufficient.

As a more realistic equation of state we consider the analytical
approximation (according to Zel'dovich and Novikov \cite{ZN})
of Tsuruta-Cameron's equation of state \cite{TC}.
In this case the interaction between the nucleons is taken
into account in a simple approximation and the pressure is given by:
\ben
p = \varepsilon + {{\rho_0 c^2} \over 2} -
{{\rho_0 c^2}\over 2}\left(1+{{4\varepsilon}\over{\rho_0 c^2}}\right)^{1/2}
\la{EOS2}
\een
where  $\rho_0 = 5* 10^{15} g/{cm}^3$ as the relation between the
particle number density and the energy density is
\ben
\varepsilon = m_{N}c^2 \left(n + m_{N}{n^2 \over \rho_0}\right)
\een

It turns out that these two examples present typical results which
qualitatively agree with the results for the other equations of state
of star's matter.

\newpage

\section{Numerical results and discussions}

We have solved the system of equations (\ref{NF}) coupled with the state
equations (\ref{EOS1}) and (\ref{EOS2}) numerically  using the method due to
Runge-Kutta-Merson with automatic error control.
The results are shown in the corresponding figures.\\
Hereafter all masses are measured in units $M_{\bigodot}$.
\vskip 2truecm

\begin{figure}[htbp]
\vspace{4truecm}
\includegraphics{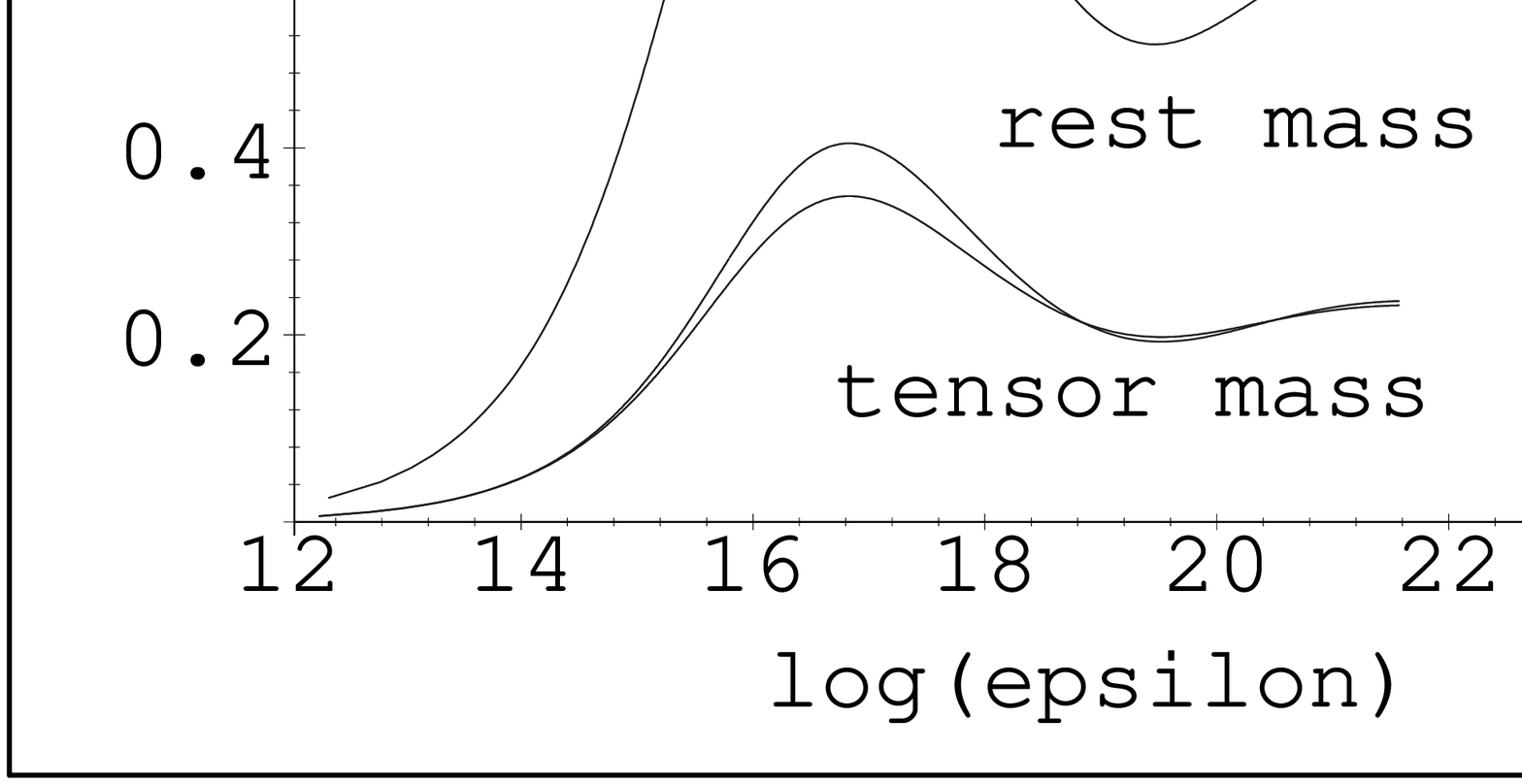}
\includegraphics{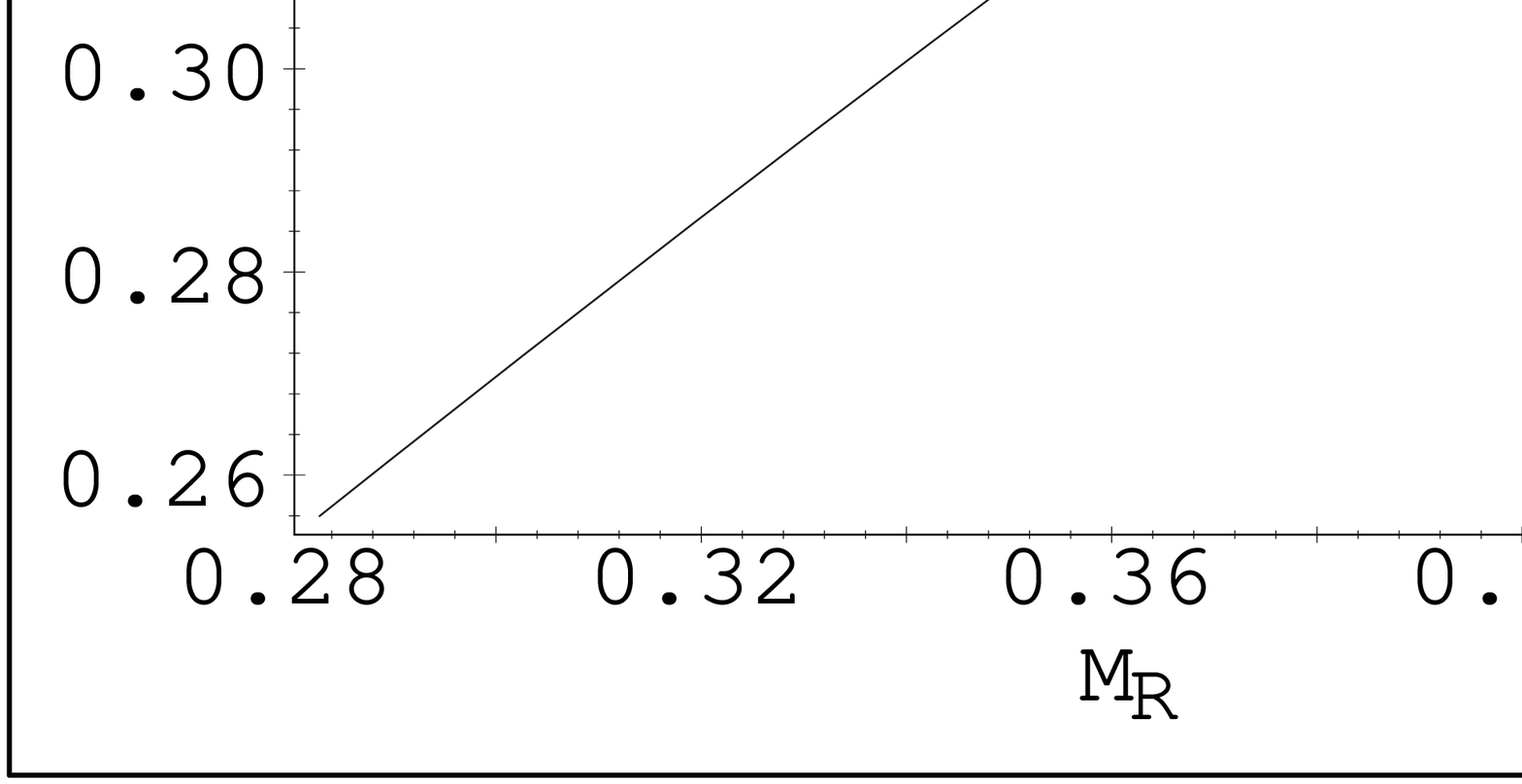}
   \vskip 0.5truecm
    \caption{a) $M -log(\varepsilon_{c})$ dependence.
    \hskip 1.8truecm b) $M_{T} - M_{R}$ dependence.\hskip .7truecm}
\vspace{.5truecm} 
    \label{Fig1}
\end{figure}
First we concentrate our attention on the case of non-interacting neutron gas.
In Fig. 1a) the dependence of the three masses $M_{T}$,$M_{K}$,$M_{R}$ on
the central density $\varepsilon_{c}$ is shown. The appearance of a cusp in
Fig. 1b) , where the dependence of the tensor mass $M_{T}$ on the rest mass
$M_{R}$ is presented, shows that their maxima lie at the same point. Although,
the maxima of the rest mass $M_{R}$ and Keplerian mass $M_{K}$ are too close
in Fig. 1a) they don't lie at the same point, as it may be seen from Fig. 2a)
which shows the dependence of $M_{K}$ on $M_{R}$. We see also that Keplerian
mass is considerably greater than the tensor one -- about three times.

In Fig. 3a) the $M_{T} - R$ dependence is represented. It's seen that the
$M_{T} - R$  curve in our case is fairly similar to the one of general
relativity, but there are significant differences, too. The maximum mass
${M_{T}}_{max} $ in our case is $ \approx 0.35M_{\bigodot}$, while in general
relativity the Oppenheimer-Volkoff's mass is $M_{OV} = 0.7M_{\bigodot}$.
The radius corresponding to the mass $M_{\bigodot}$ is  $R = 4.2 km $,
while in the case of general relativity  $ R = 9.6 km $.
If we look at Fig. 2b) where the dependence
of $M_{T}$ on the central density ${\varepsilon}_{c}$ is shown,
we note that ${M_{T}}_{max}$ lies at
${\varepsilon}_{c} \approx 4.5 * 10^{16} g/{cm}^3$, while  $M_{OV}$ lies at
${\varepsilon}_{c} \approx  5 * 10^{15} g/{cm}^3$ in general relativity.
The average  density in our case is about $ 4 $ times greater than the one
in general relativity. Hence, in the model under consideration the neutron
star is more compact and has a mass about $1/2$ - times smaller than $M_{OV}$.
In Fig. 3b) the dependence of Keplerian  mass on the star radius is
presented. It's seen that the Keplerian mass is about $1.5$ times greater than
$M_{OV}$.
\vskip 2truecm

\begin{figure}[htbp]
\vspace{4truecm}
\includegraphics{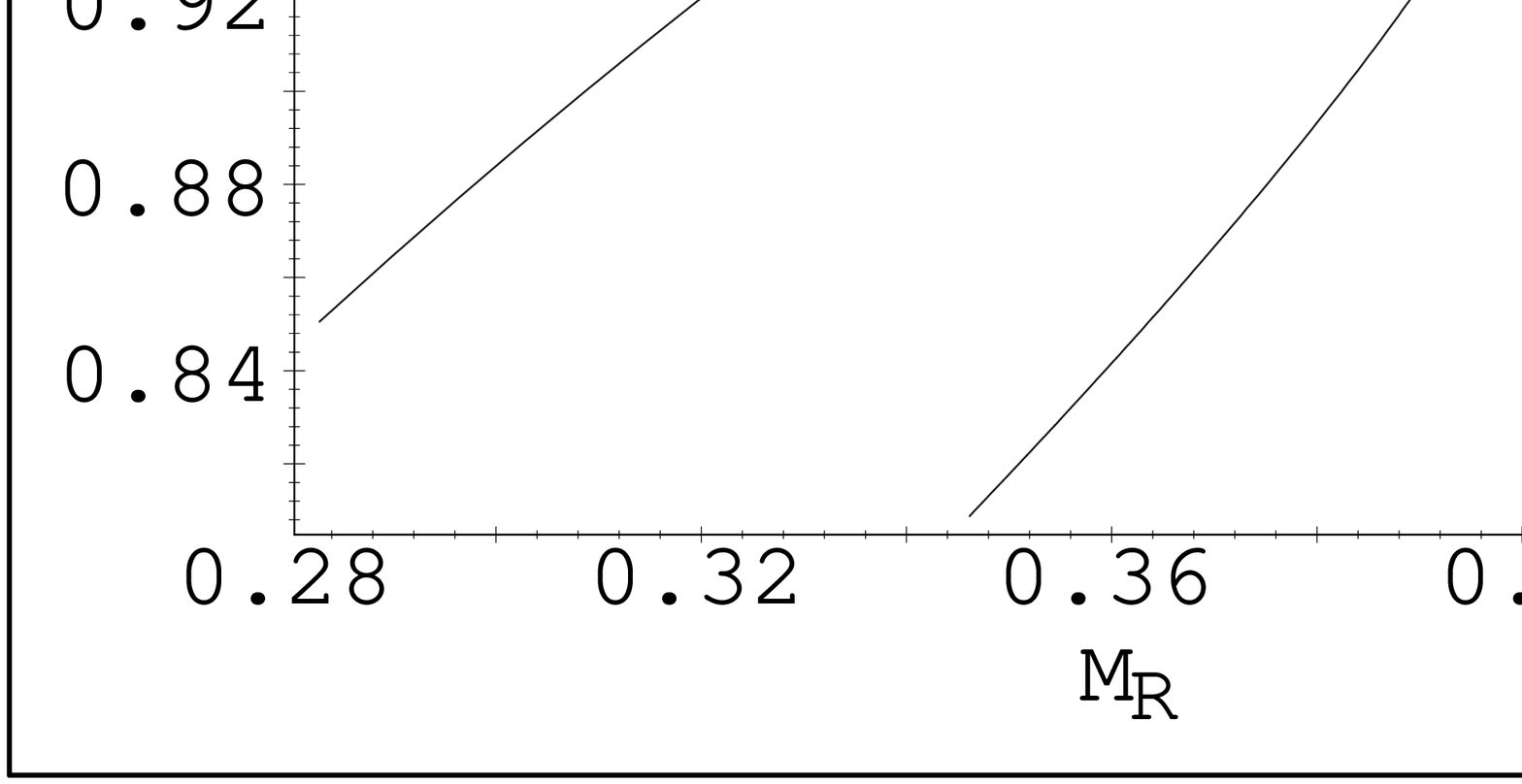}
\includegraphics{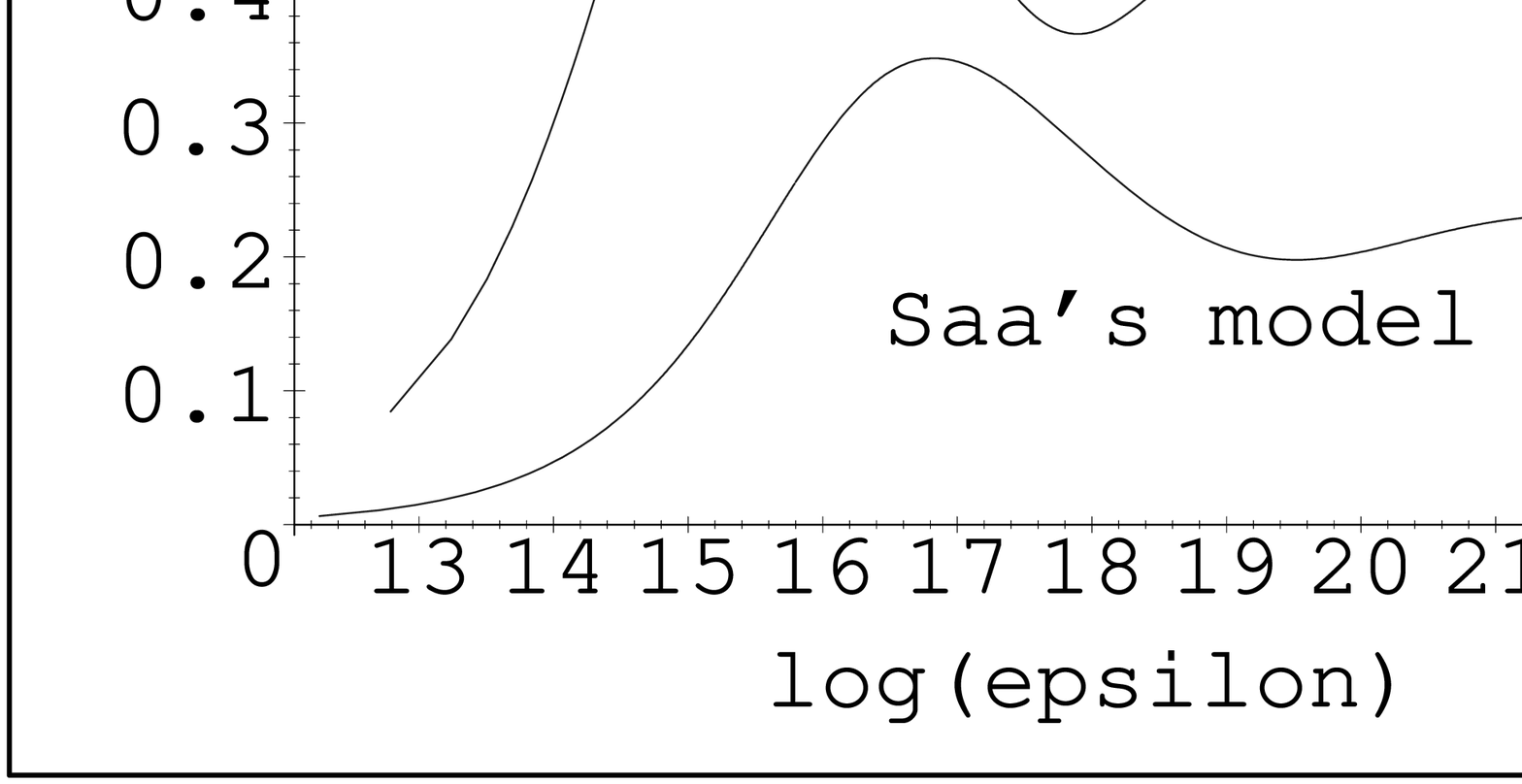}
    \vskip 0.5truecm
    \caption{a) $M_{Kepler}-M_{R}$ dependence.
    \hskip 2truecm b) $M_{T} -log(\varepsilon_{c})$ dependence.\hskip .5truecm}
\vspace{.5truecm} 
    \label{Fig2}
\end{figure}

\vskip 2truecm

\begin{figure}[htbp]
\vspace{4truecm}
\includegraphics{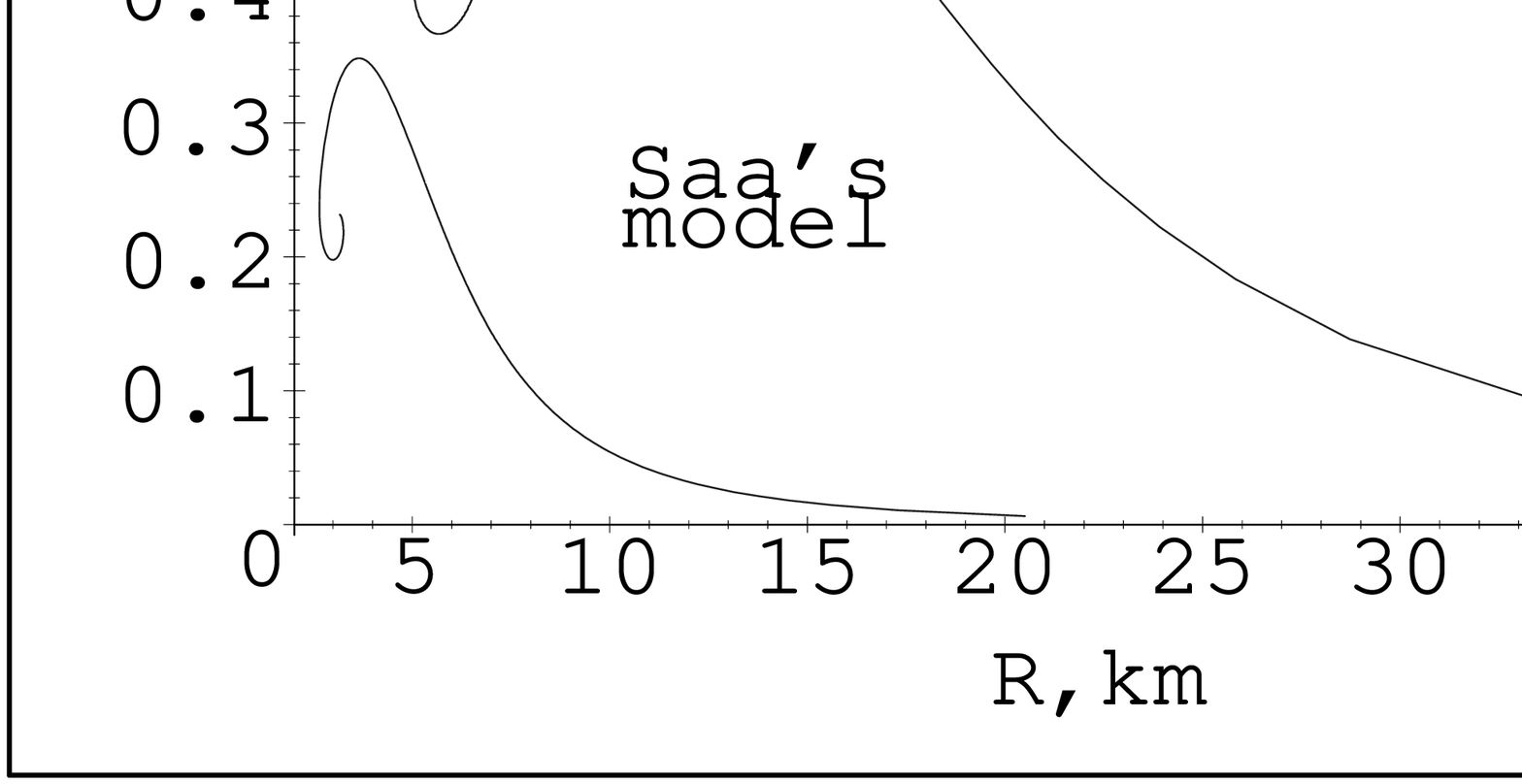}
\includegraphics{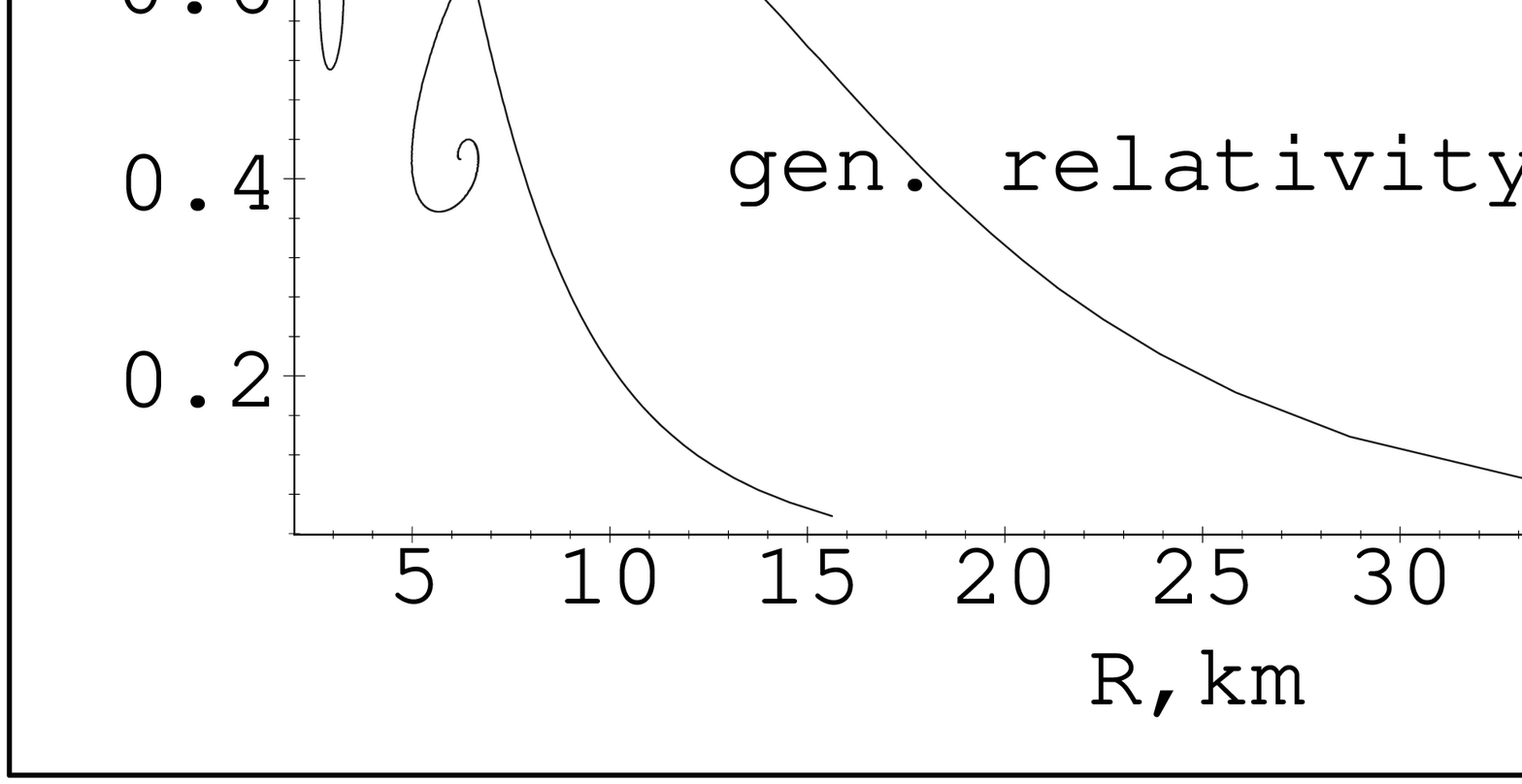}
    \vskip 0.5truecm
    \caption{a) $M_{T}-R$ dependence.
    \hskip 2.4truecm b) $M_{Kepler}-R$ dependence.\hskip  1.1truecm }
\vspace{.5truecm}
 \label{Fig3}
\end{figure}

\vskip 2truecm

\begin{figure}[htbp]
\vspace{4truecm}
\includegraphics{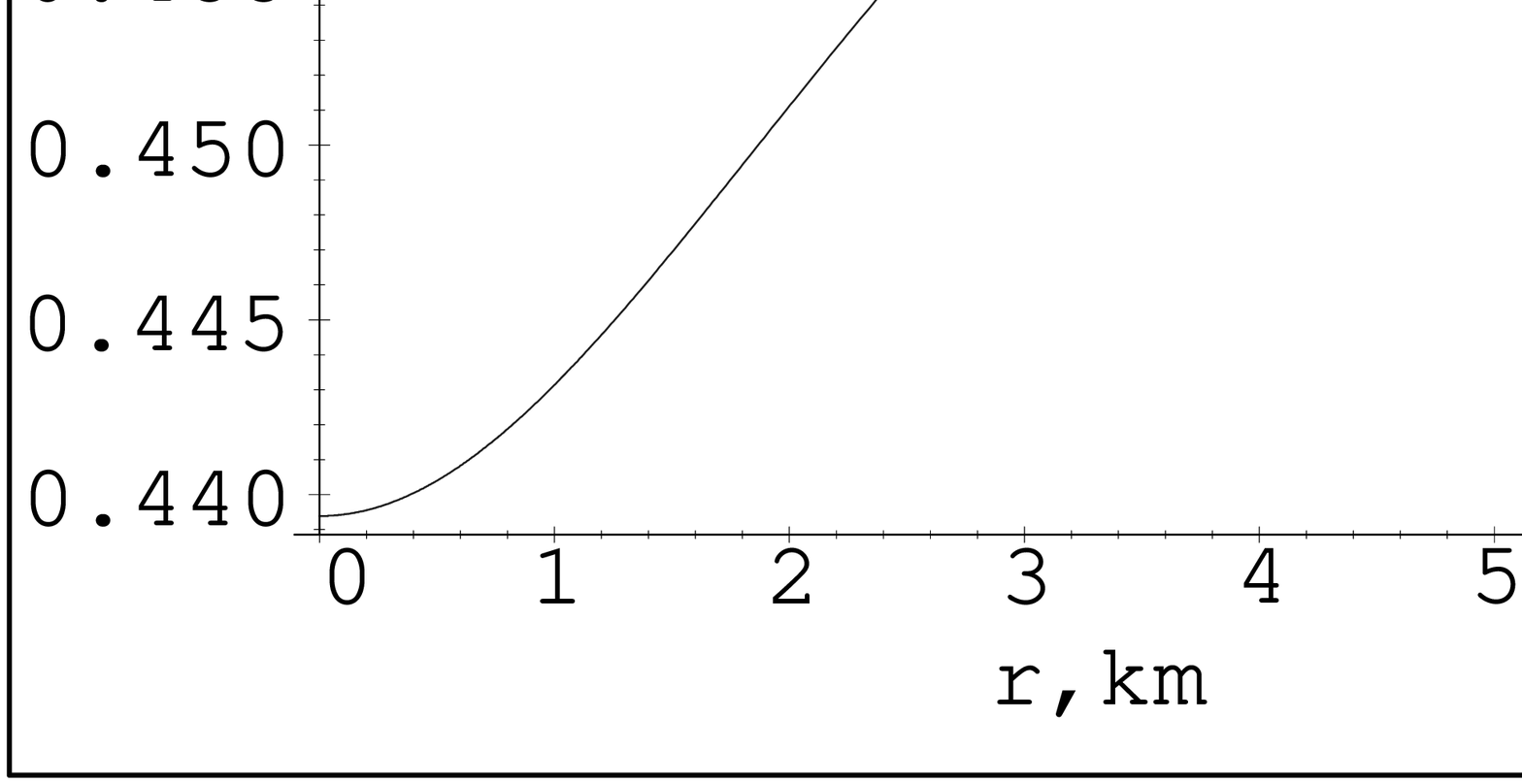}
\includegraphics{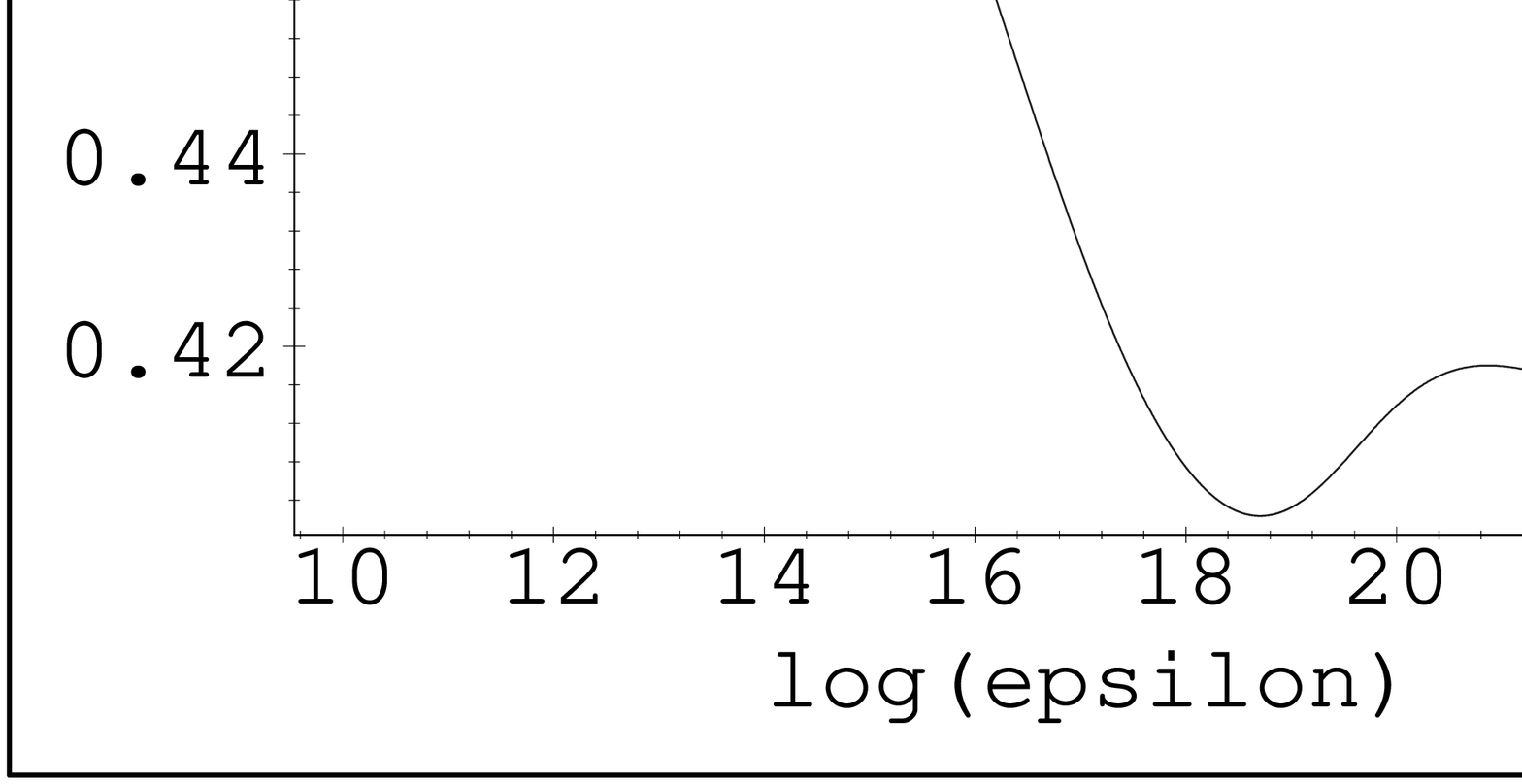}
    \vskip 0.5truecm
    \caption{a) $k- r$ dependence.
    \hskip 2truecm b) $K -log(\varepsilon_{c})$ dependence.\hskip  .5truecm}
\vspace{.5truecm} 
    \label{Fig2_}
\end{figure}

In the  Fig. 4a) the dependence $k(r)$ is shown inside the star
(for central density $7.5*10^{15}g/{cm}^3 $).
In accordance to
the general considerations $k$ increases from the center of the star to the
surface, where $k$ takes a value $K=k(R) \approx 0.45 - 0.46$,
which is close to $0.5$.
The dependence $K(\varepsilon_{c})$ of $K$ on the star central density
$\varepsilon_{c}$ is shown in Fig. 4b).
It's seen that $K$ decreases when density increases, which is similar to the
previous case. So, the ratio of the torsion force to the gravitational one
takes its minimum value at the center of the star and is the greatest at the
surface.
\vskip 2truecm

\begin{figure}[htbp]
\vspace{4truecm}
\includegraphics{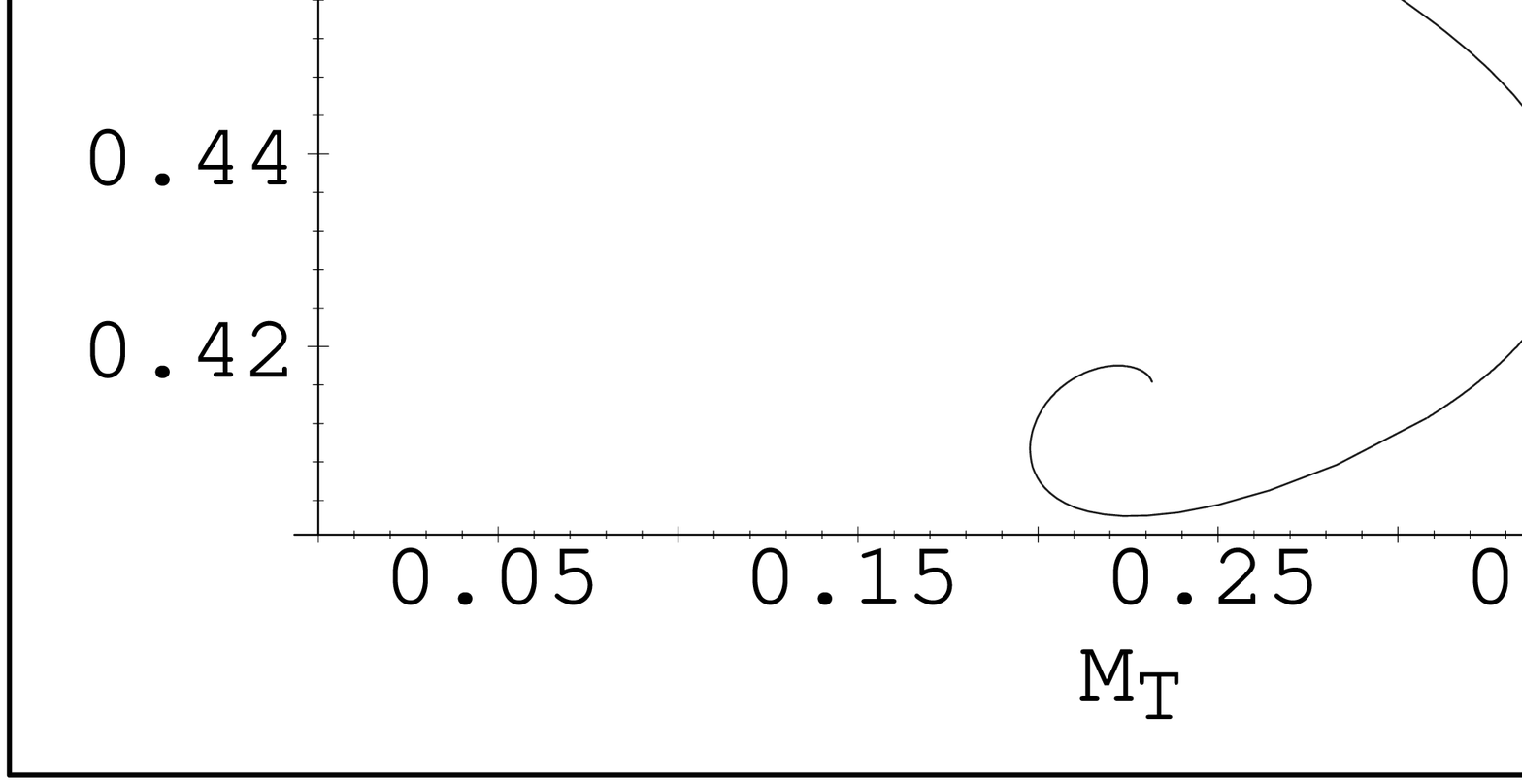}
\includegraphics{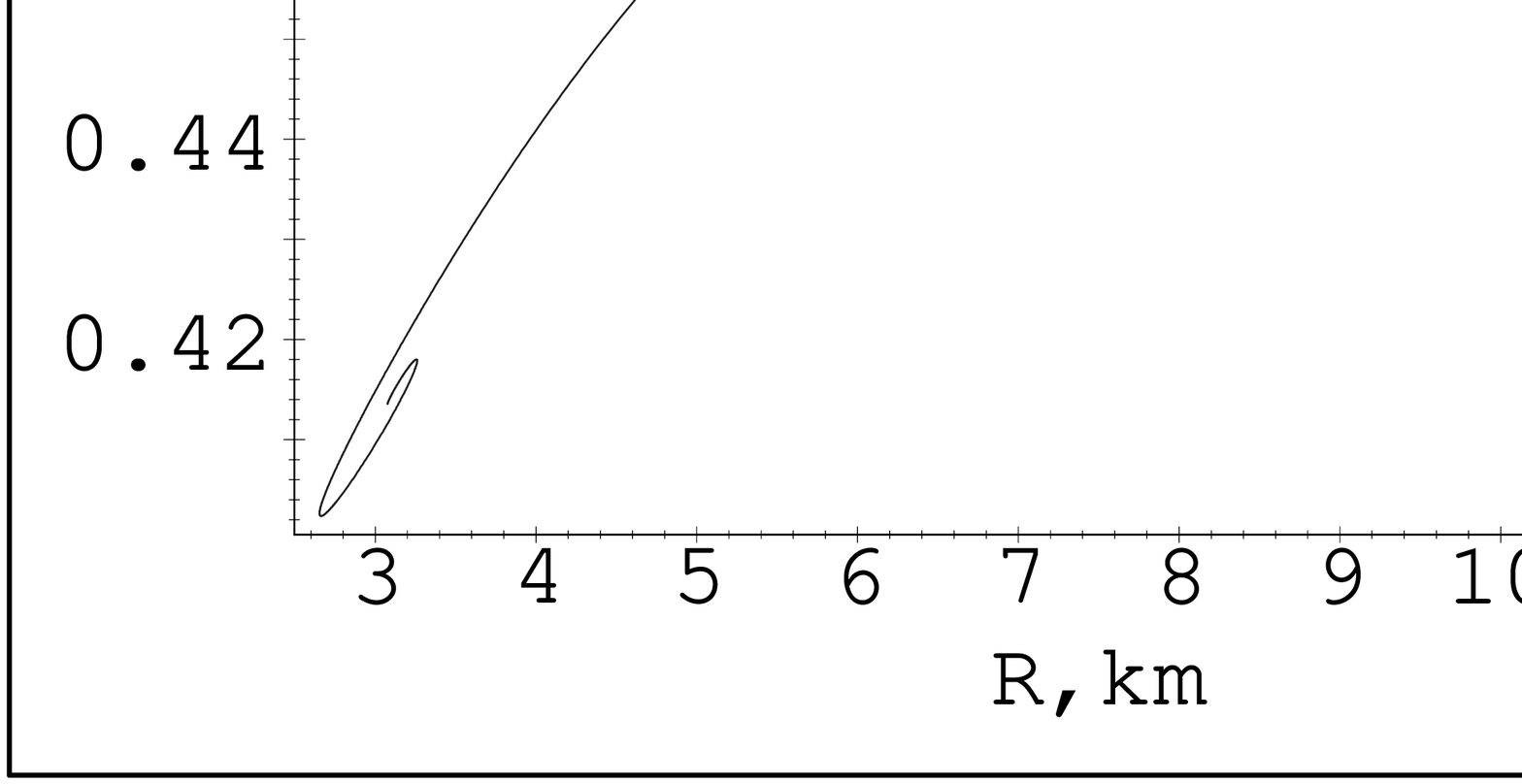}
    \vskip 0.5truecm
    \caption{a) $K-M_{T}$ dependence.
    \hskip 2.4truecm b) $K-R$ dependence.\hskip  1.1truecm }
\vspace{.5truecm}
 \label{Fig3_}
\end{figure}

 As it may be seen from Fig. 5a) expressing the dependence $K(M_{T})$, the
torsion-urged effects are relatively strongest in the case of small
masses -- with increasing of the star mass
(up to the point where the star loses its
stability) $K$ decreases.
It's seen from Fig. 5b), where the dependence of $K$
on the star radius $R$ is shown, that in the area of stability  $K$ decreases
when $R$ decreases too -- the more compact stars are, the smaller $K$ they
have.

Fig. 6a) presents the dependencies $\theta(r)$ and $\nu(r)$ inside the star.
One may see that ${1 \over 2}\nu - 3\Theta < 0$  everywhere.
The dependencies $m_{T}(r)$,$m_{Kepler}(r)$ and $m_{\theta}(r)$ are shown
in Fig. 6b)  for central density $7.5*10^{15}$. As it has already mentioned
all masses  increase with $r$.
\vskip 2truecm

\begin{figure}[htbp]
\vspace{4truecm}
\includegraphics{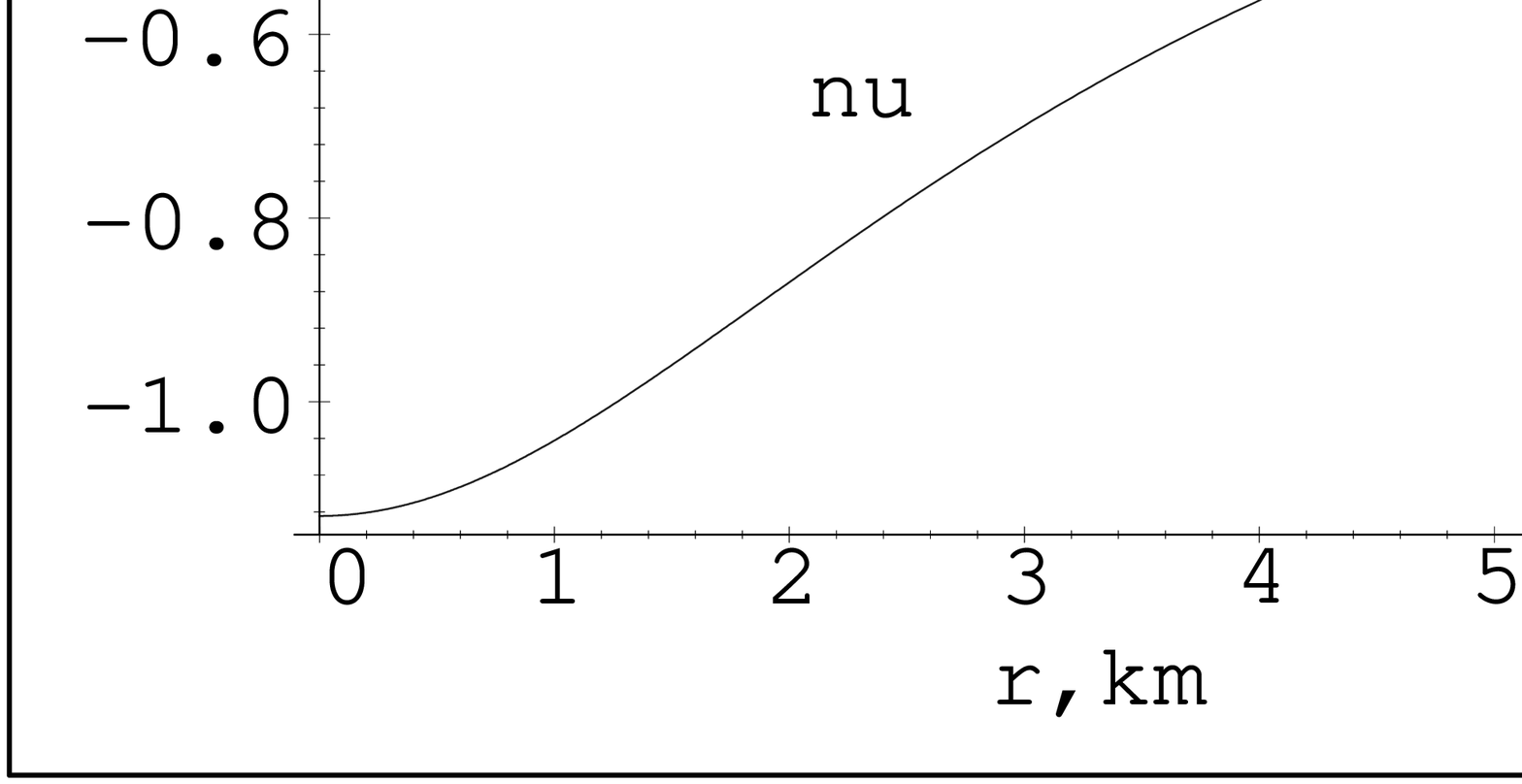}
\includegraphics{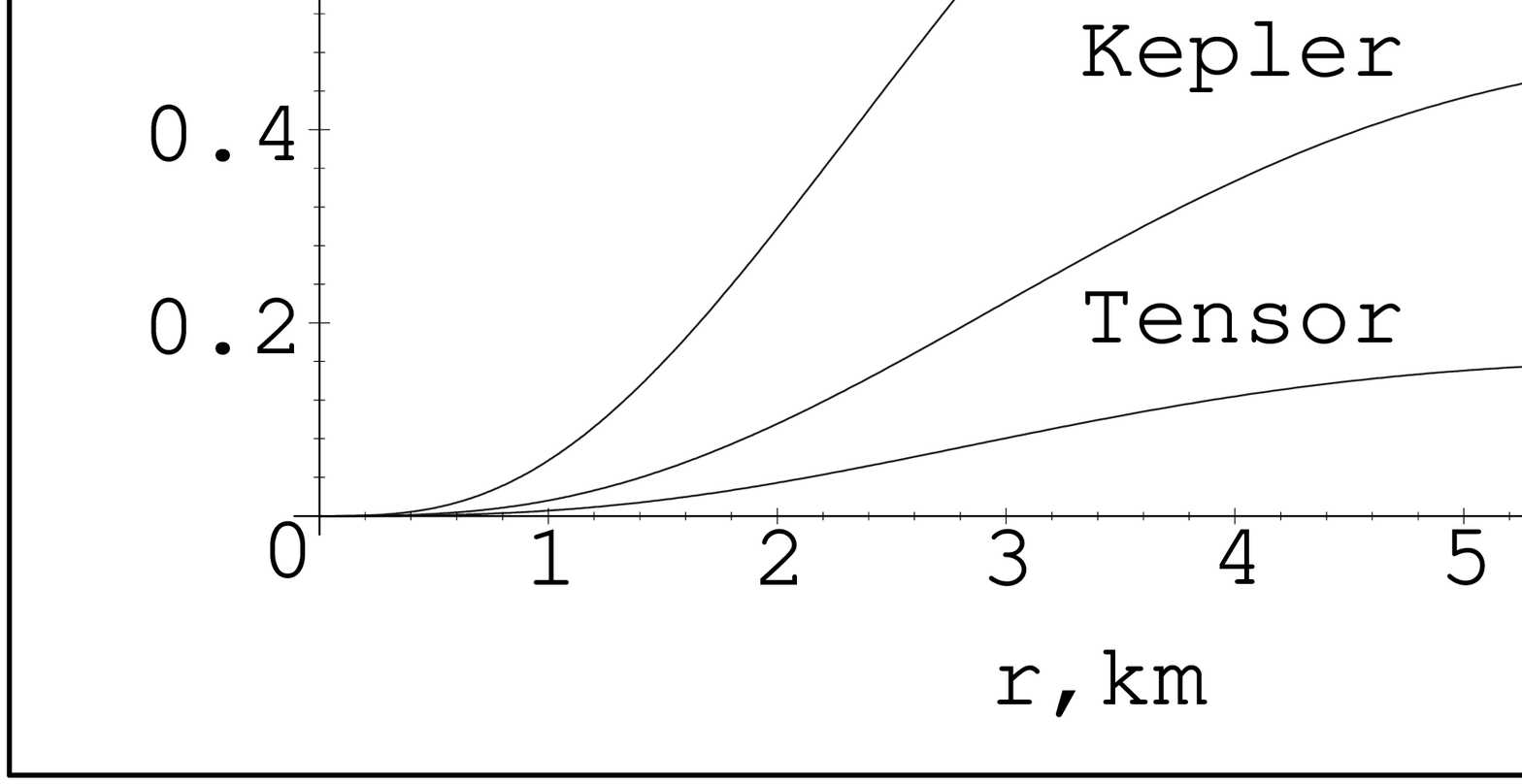}
    \vskip 0.5truecm
    \caption{a) $\Theta, \nu -r$ dependence.
\hskip 1.9truecm b)$m_{T},m_{Kepler},m_{\theta} -r$ dependence.\hskip .7truecm }
\vspace{.5truecm} 
    \label{Fig4}
\end{figure}

The following figures illustrate the case of Tsuruta-Cameron equation
of state (TCES).
\vskip 2truecm

\begin{figure}[htbp]
\vspace{4truecm}
\includegraphics{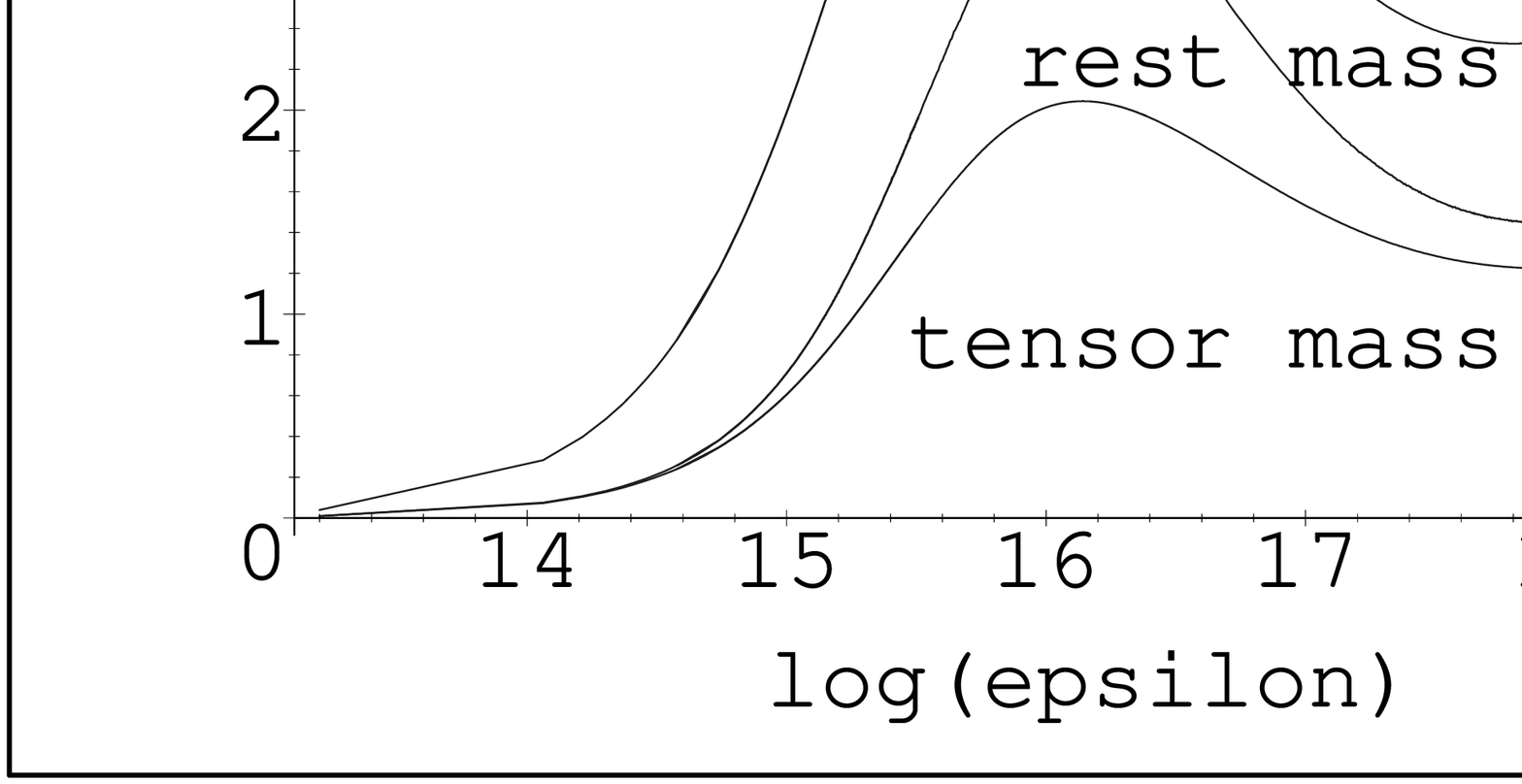}
\includegraphics{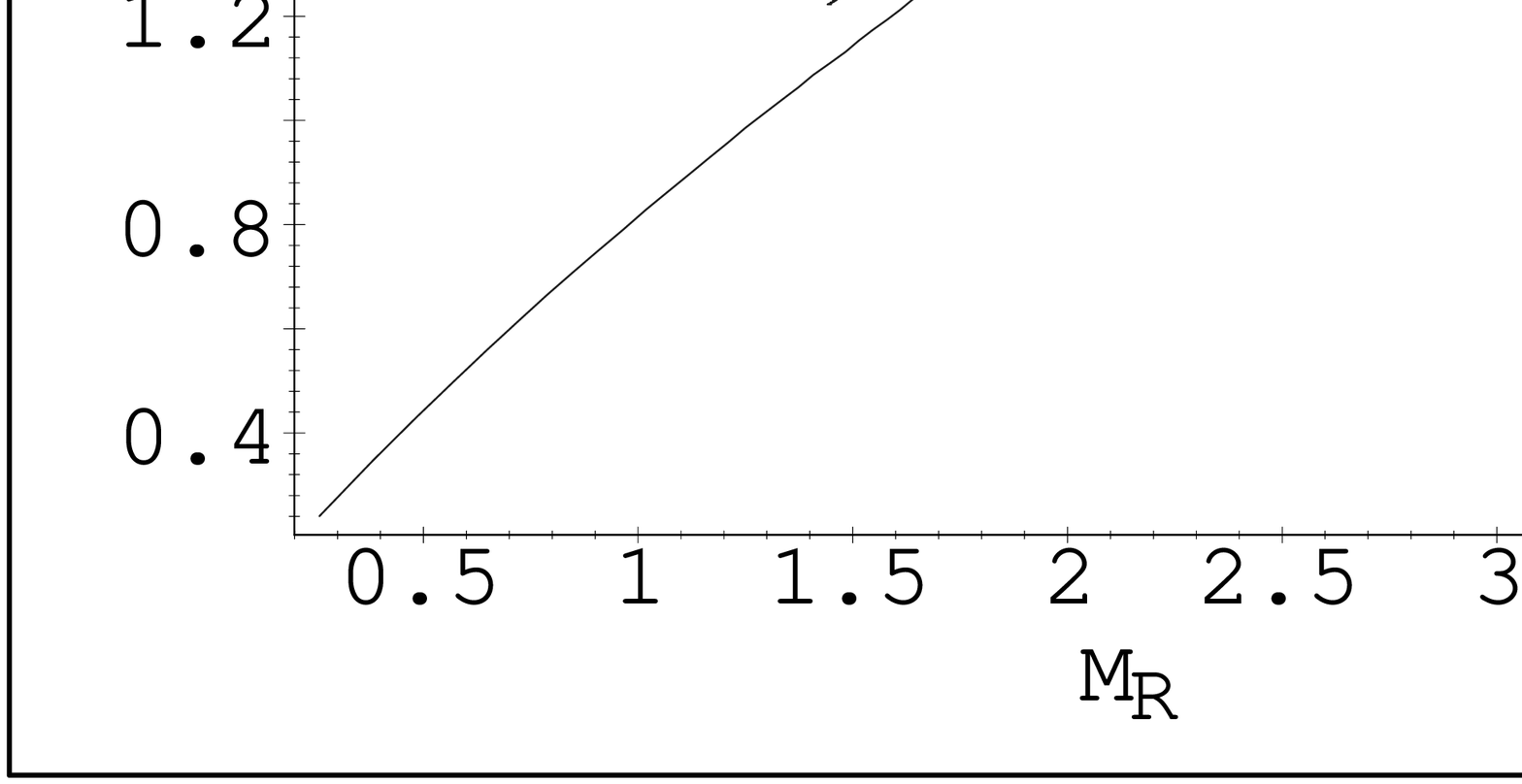}
    \vskip 0.5truecm
    \caption{TCES. a) $M -log(\varepsilon_{c})$ dependence .
    \hskip 1.2truecm b) $M_{T} -M_{R}$ dependence.\hskip  1.2truecm }
\vspace{.5truecm} 
    \label{Fig5}
\end{figure}

We see from the figures that the maximum tensor mass in this case is
about $2M_{\bigodot}$ and the corresponding radius is about $7.5km$ - the
same quantities in general relativity are correspondingly
$\approx 1.6M_{\bigodot}$ and $\approx 11.5km$.
Hence, the interaction between the nucleons leads to an increase
in the maximum mass, as in general relativity.

Note the differences between the Fig. 5b) and Fig. 6b)
(for the case of non-interacting neutron gas),
and the corresponding  Fig. 11b) and Fig. 12b)
(for the case of Tsuruta-Cameron equation of state).
There one can see the strong dependence of some results
in the Saa's model of gravity with propagating torsion
on the equation of state of star's matter.

\vskip 2truecm

\begin{figure}[htbp]
\vspace{3.5truecm}
\includegraphics{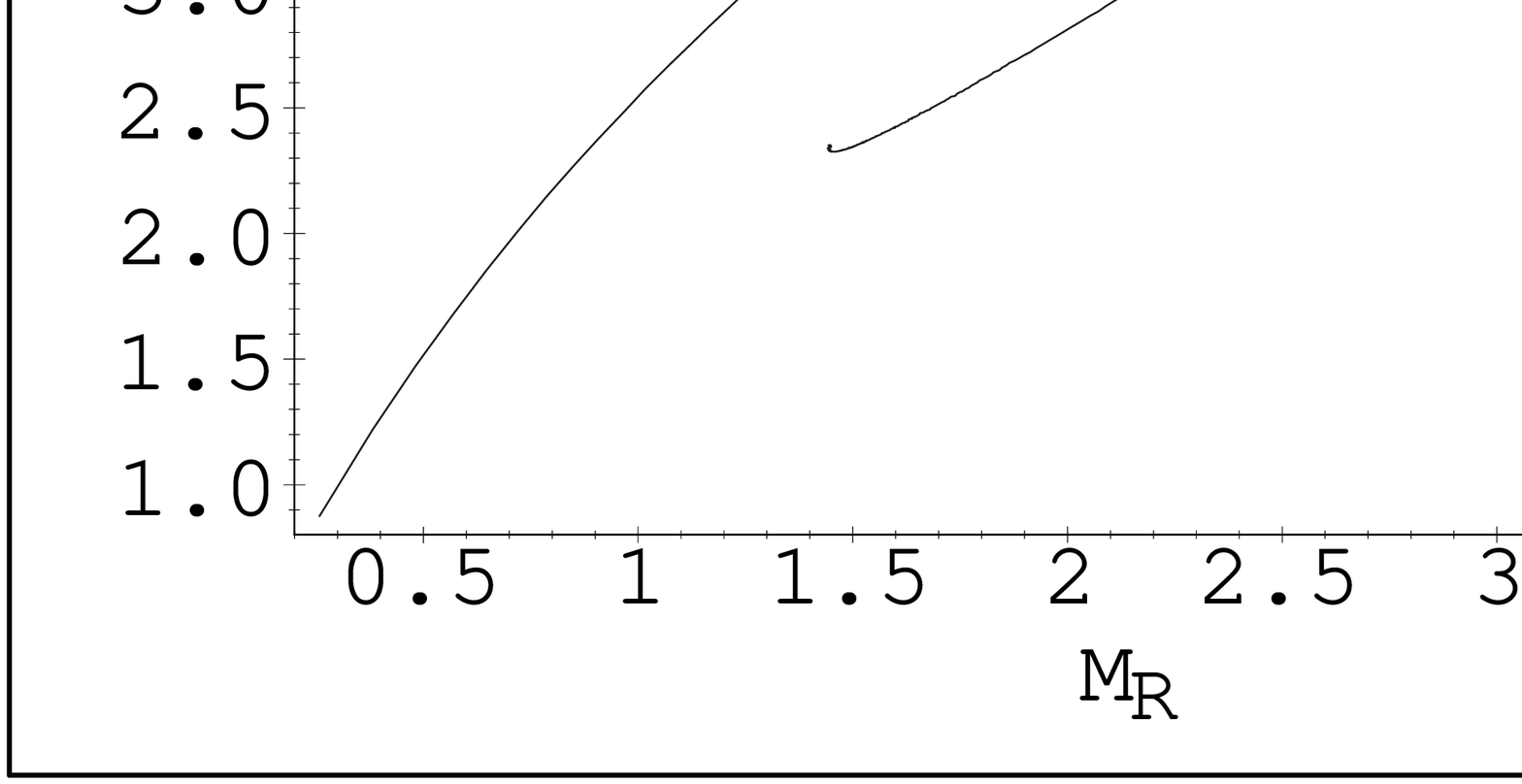}
\includegraphics{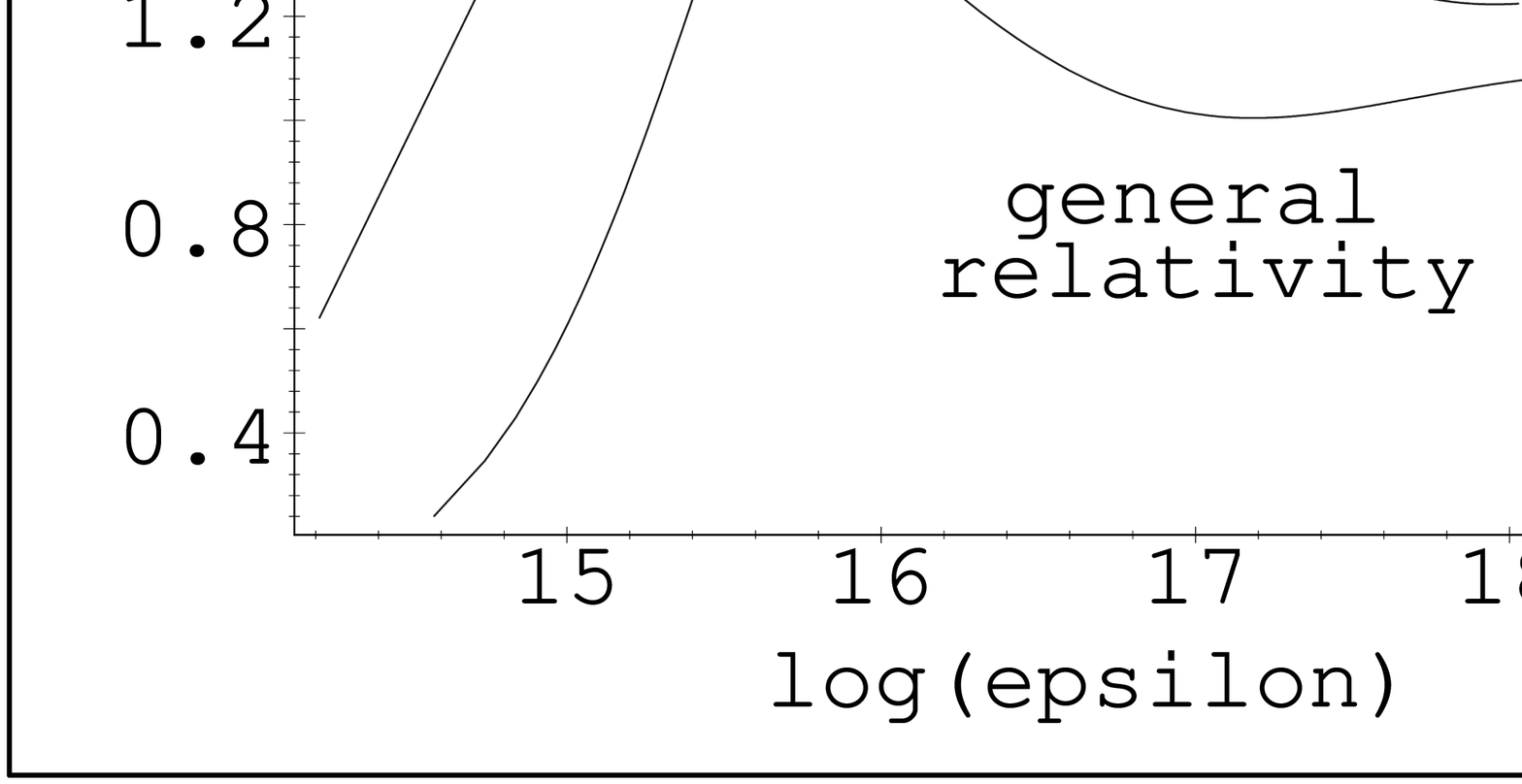}
    \vskip 0.5truecm
    \caption{TCES. a)$M_{Kepler}-M_{R}$ dependence.
    \hskip .4truecm b) $M_{T}-log(\varepsilon_{c}) $ dependence.\hskip 1truecm}
\vspace{.5truecm} 
    \label{Fig6}
\end{figure}

\begin{figure}[htbp]
\vspace{4truecm}
\includegraphics{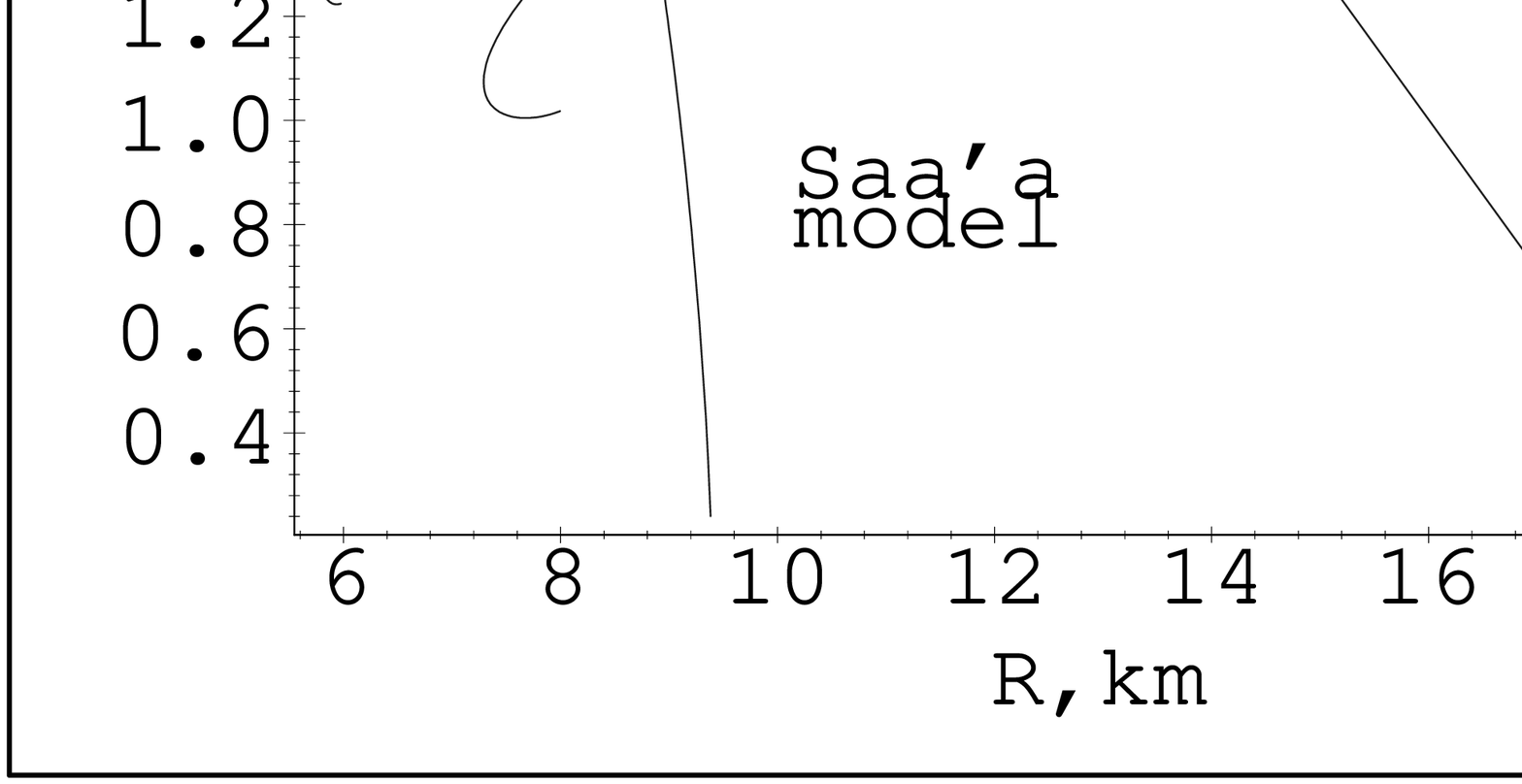}
\includegraphics{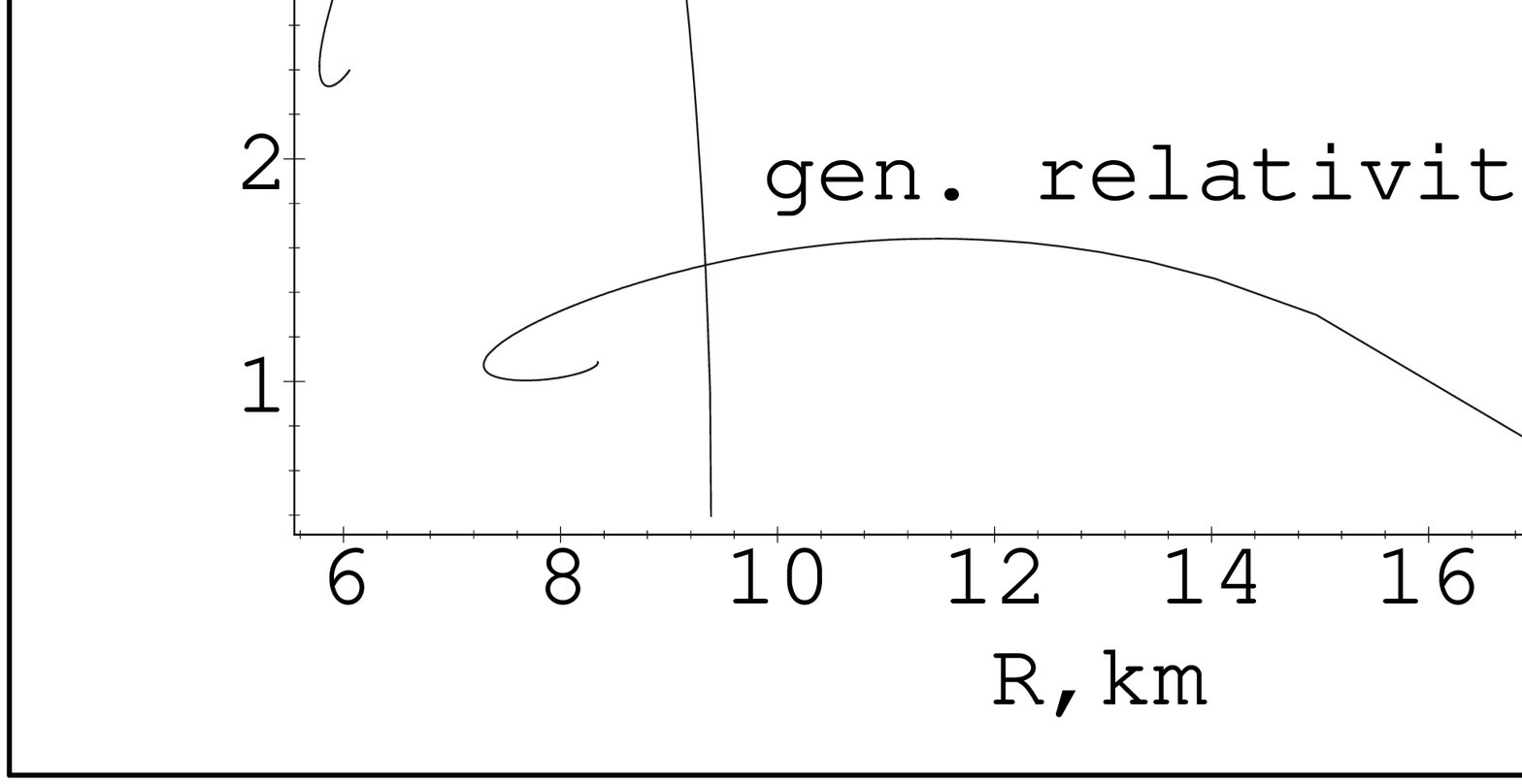}
    \vskip 0.5truecm
    \caption{TCES. a) $M_{T}-R$ dependence.
    \hskip 1.6truecm b) $M_{Kepler}-R$ dependence.\hskip  1.7truecm}
\vspace{.5truecm} 
    \label{Fig7}
\end{figure}


\begin{figure}[htbp]
\vspace{4truecm}
\includegraphics{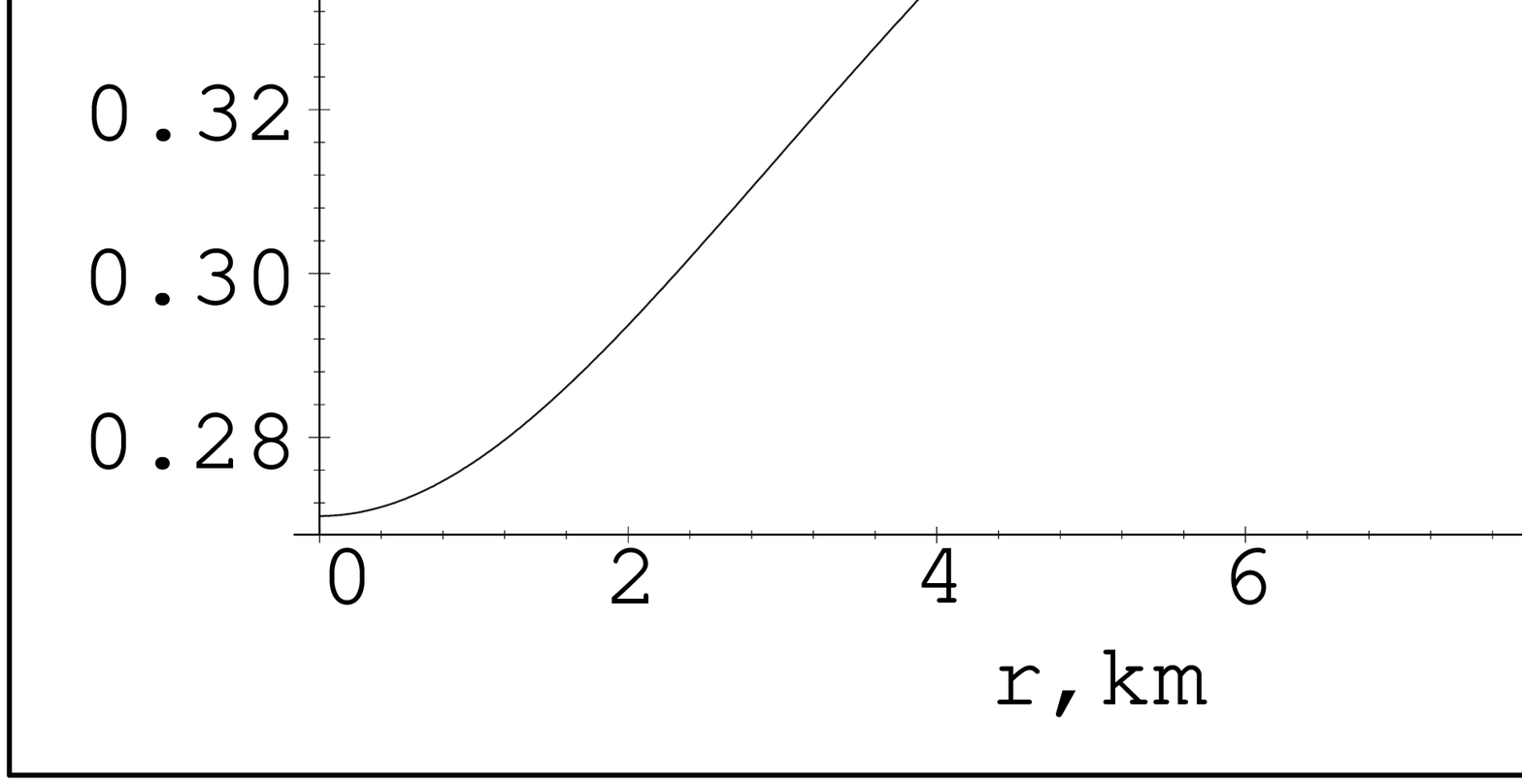}
\includegraphics{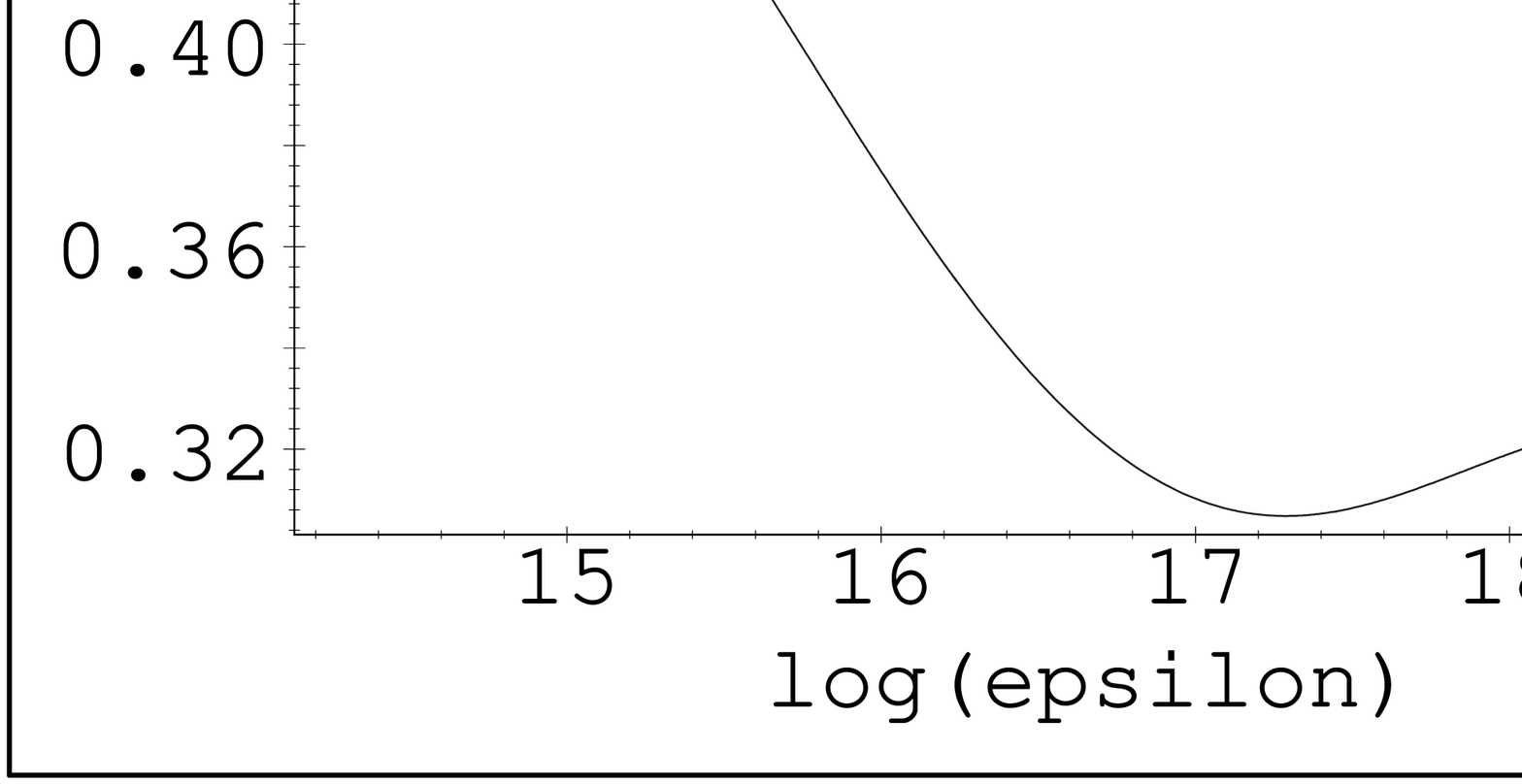}
    \vskip 0.5truecm
    \caption{TCES. a) $k-r$ dependence.
    \hskip 1.3truecm b) $K -log(\varepsilon_{c}) $ dependence.\hskip 3truecm}
\vspace{.5truecm} 
    \label{Fig8}
\end{figure}

\begin{figure}[htbp]
\vspace{4truecm}
\includegraphics{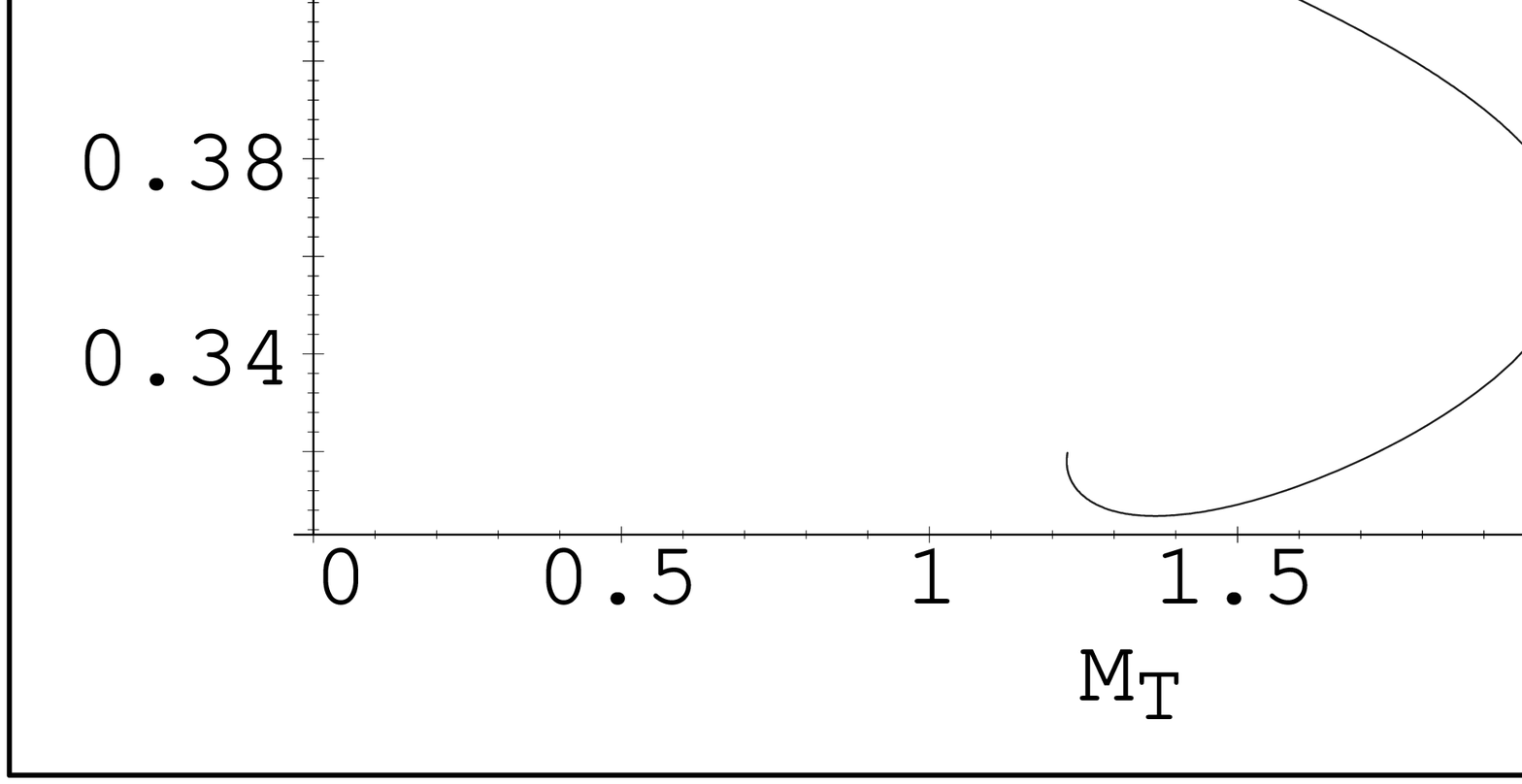}
\includegraphics{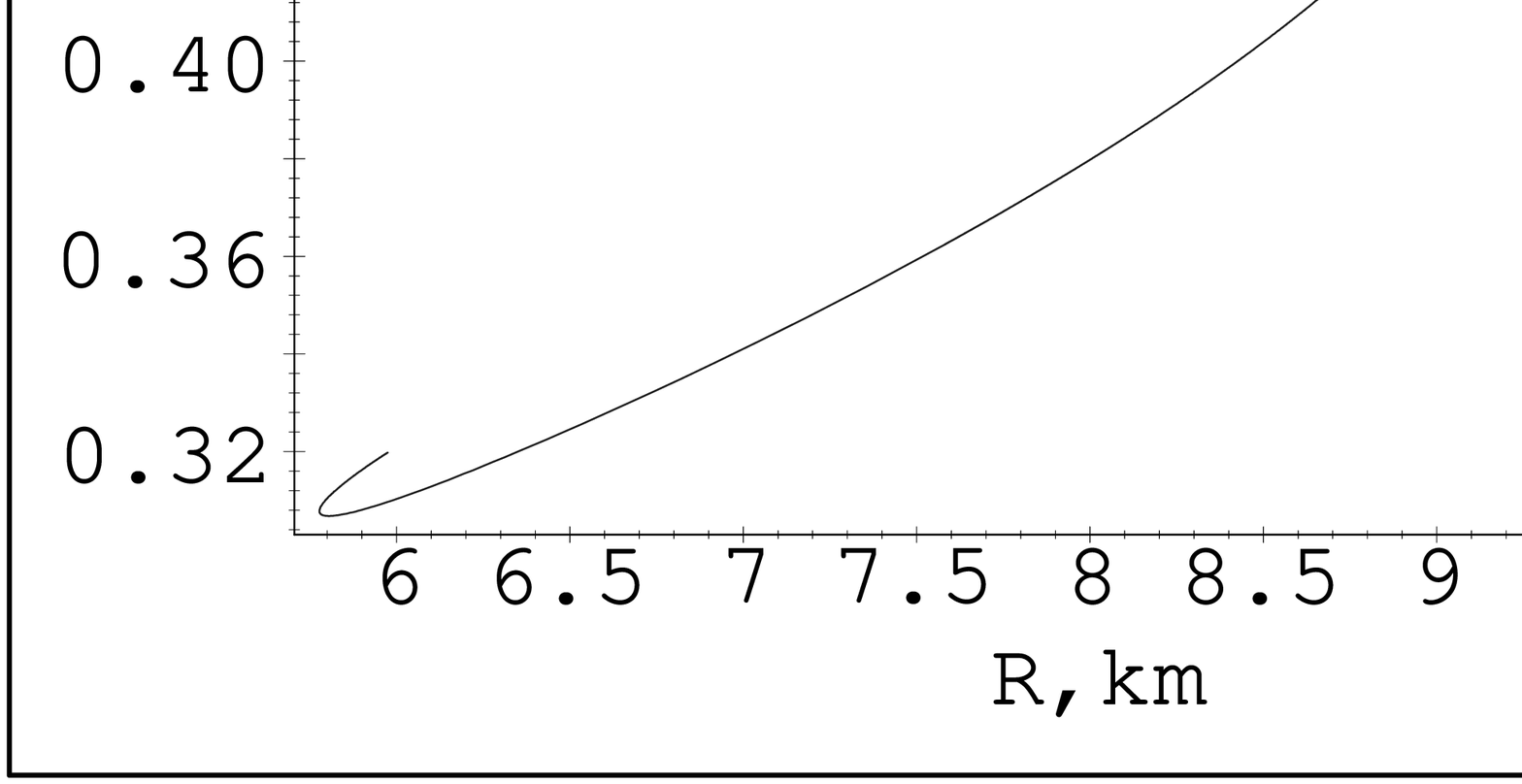}
    \vskip 0.5truecm
    \caption{TCES. a) $K-M_{T}$ dependence.
    \hskip 1.3truecm b) $K-R $ dependence.\hskip 3truecm}
\vspace{.5truecm} 
    \label{Fig8_}
\end{figure}

\begin{figure}[htbp]
\vspace{4truecm}
\includegraphics{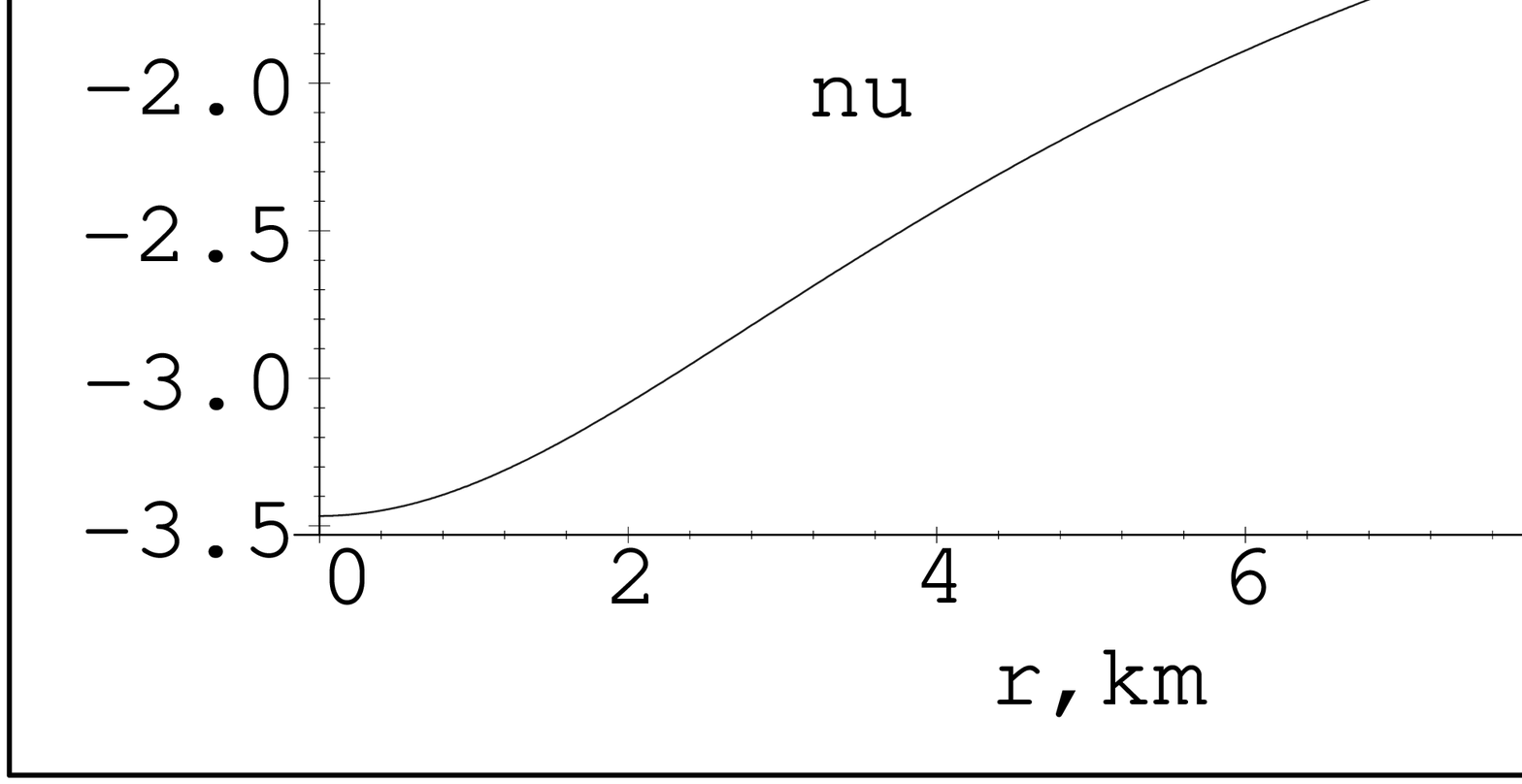}
\includegraphics{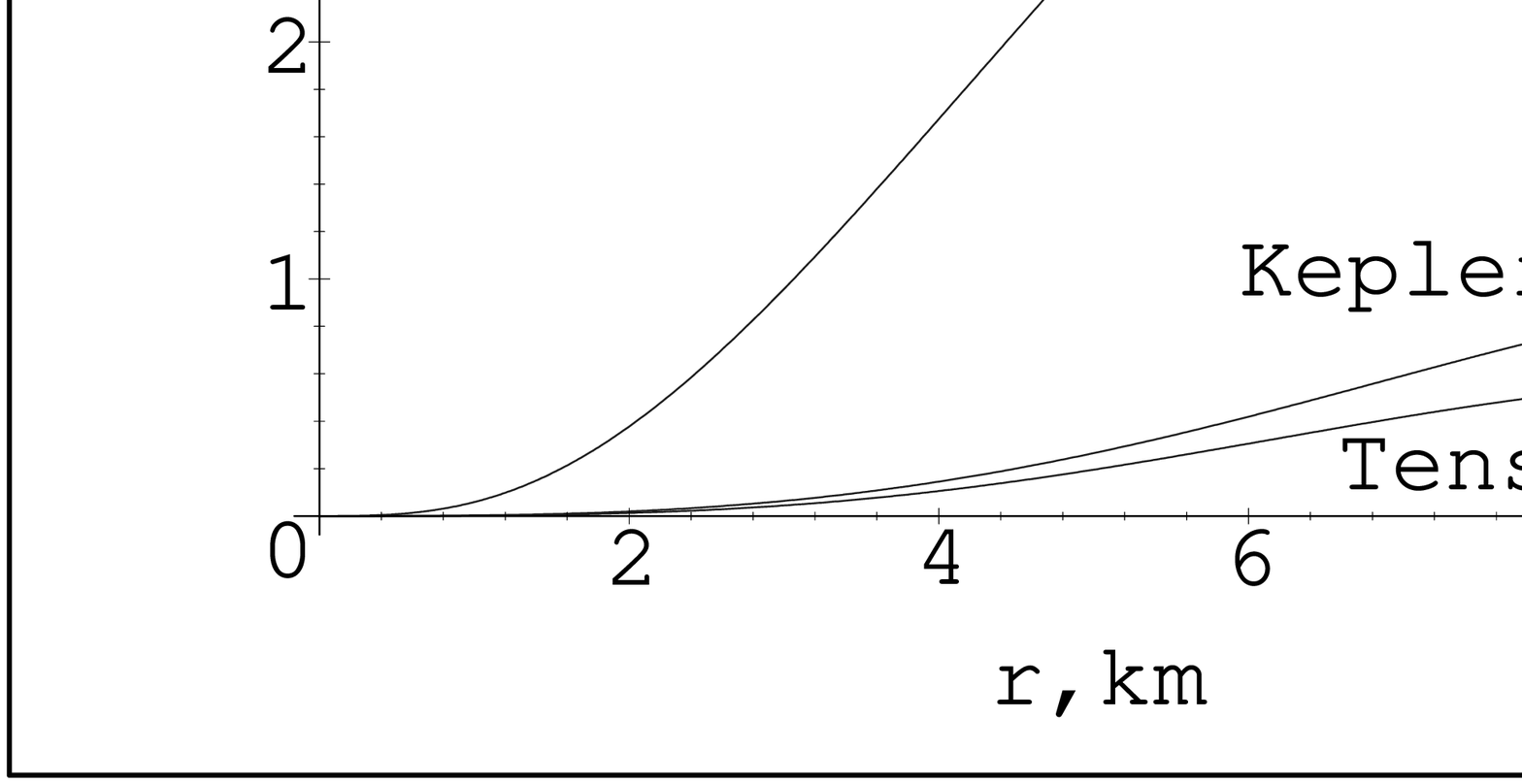}
    \vskip 0.5truecm
    \caption{TCES. a) $\Theta, \nu-r$ dependence.
\hskip 1.3truecm b) $m_{T},m_{Kepler}, m_{\theta}-r$ dependence.\hskip 3truecm}
\vspace{.5truecm} 
    \label{Fig8+}
\end{figure}
We have also examined the Harrison-Wheeler's equation of state \cite{HTWW}.
As in general relativity the numerical results are very close to these
for the noninteracting neutron gas.
For example the maximum tensor and  Keplerian mass
is correspondingly $\approx 0.35M_{\bigodot}$ and $\approx 1M_{\bigodot}$,
and the corresponding radius is $3.8 km$.

Other equations of state (of politropic type) have been examined, too.
The corresponding maximum tensor mass of a neutron star
reaches a value about $2.5-2.6M_{\bigodot}$, while the corresponding
Keplerian mass is about $6-6.5M_{\bigodot}$.

As it is seen from numerical calculations the tensor mass and Keplerian mass
differ very significantly from each other. This behaviour of the  masses
is qualitatively the same as in the case of a boson star in Brans-Dicke
theory with $\omega=-1$  \cite{Whinnett}.
Saa's model corresponds to the value of the Brans-Dicke parameter
$\omega=-{4\over 3}$ which is close to $-1$. That's why the observed
qualitative agreement is natural.  Hence, including a scalar field with
an approximately the same $\omega$ in physically different kinds of
stars  we find similar departures of the corresponding predictions of
general relativity.

This conclusion holds also in the case of large $\omega$. For example,
in \cite{TLS} and \cite{TSL} Brans-Dicke boson star has been
considered with $\omega = 400$. There the results are quite similar to
corresponding ones in general relativity which is just the limit
$\omega \to \infty$.
The star is a bit lighter but has a higher density in future.

\section{Stability analysis via catastrophe theory}

Methods of catastrophe theory have recently been applied for investigation
the stability of self-gravitating systems as neutron and boson stars in
\cite{KMS1} and \cite{KMS2} (in the case of boson stars see also \cite{CS}
and \cite{TSL}).

Here we discuss briefly the important question of the stability of a
neutron star in our case using tools of catastrophe theory.

The basis for the stability analysis are figures Fig.1b and Fig.7b.
The conserved quantities which we use to calculate the binding energy are
the tensor mass and the rest mass (particle number). Drawing them against
each other we obtain so-called bifurcation diagram. Actually, Fig.1b and
Fig.7b are bifurcation diagrams. When the central density increases
one meets a cusp. The appearance of a cusp itself isn't enough
to conclude that the stability of the neutron star changes.
However, if it's known that the star is stable along the first brunch
then it will become unstable along the second brunch.
In general relativity the stability  of the neutron stars for small
central densities is proved by using linear perturbation analysis.
As it's seen from the figures, the behaviour of the basic conserved
quantities in our case is qualitatively the same  as
in general relativity. That's why it's natural to assume that for small
central densities, i.e. along the first brunch the star in our case is
stable against small radial perturbations. Besides of that the first brunch
is composed completely of negative binding energy states which shows that
it is potentially stable.

Making use of the above considerations we can conclude that the neutron
stars with central densities from small values to the cusp are stable.
Beyond the cusp one radial perturbation mode develops instability and
the star becomes unstable.

\newpage

\section{Saa's model, Roll-Krotkov-Dicke,
Braginsky-Panov and E\"otvosh experiments}

Despite the obvious beauty of Saa's model we shall show that it contradicts
to the experiments by Roll-Krotkov-Dicke, Braginsky-Panov and E\"otvosh.
To do this we need the equation of motion of an isolated test body
corresponding to the model under consideration. In our case the local
conservation law for energy-momentum tensor can be written in the form
(see (\ref{DivT}) and  (\ref{SCurv}))
\ben
{\stackrel{ \{\} } {\nabla} }{}_\sigma (e^{-3\Theta} T_\alpha^\sigma) =
3e^{-3\Theta} \left({\cal L}_M -
{\sfrac 1 3}{\delta{\cal L}_M \over \delta \Theta}\right)S_\alpha.
\een
Defining the four-momentum and the mass-center of a small test body as
$$P^{\alpha}=\int T^{\alpha 0}\sqrt{|g|}e^{-3\Theta}d^3x,$$
$$X^{\beta}={\int x^{\beta}T^{0 0}\sqrt{|g|}e^{-3\Theta}d^3x \over P^{0}}$$
and following the same procedure as in general relativity
\cite{Synge}, \cite{Ni} we obtain the
equation of motion of such body in our case:\ben
P^{0}{d^2X^{\beta}\over d{x^0}^2} +
{\stackrel{\{\}}{\Gamma}}_{\mu\nu}{}^{\beta}(X)E^{\mu\nu}=
\left({\stackrel{\{\}}{\Gamma}}_{\mu\nu}{}^{0}(X)E^{\mu\nu}
\right){dX^{\beta}\over dx^0}  +
S^{\beta}(X)\Lambda  -
S^{0}(X)\Lambda{dX^{\beta}\over dx^0}
\een
where
$$E^{\mu\nu}= \int T^{\mu\nu}\sqrt{|g|}e^{-3\Theta}d^3x $$   and
$$\Lambda = 3\int\left({\cal L}_M -
{\sfrac 1 3}{\delta{\cal L}_M \over \delta \Theta}\right)
\sqrt{|g|}e^{-3\Theta}d^3x .$$

We consider a static spherically-symmetric case and therefore we put
$S^{0}=0$. Fixing a local inertial frame in which the test body is at rest
at the moment considered and the Christoffel symbols vanish at the location
of body we have
\ben
M{d^2X^{\beta}\over dt^2} =S^{\beta}(X)\Lambda
\la{TBEM}
\een
where $M$ is the inertial mass of the small test body.

Therefore in the model under consideration test bodies with different
$\Lambda$ would accelerate differently in the solar field (or in the
earth's field).
In good approximation we can consider the test body as built by
electromagnetically interacting particles in gravitational and torsion
field, i.e. the Lagrangian density for the test body is taken in the form
\ben
{\cal L}_M = {\cal L}^{em}_M  +  {\cal L}^{pf}_M
\een
where $${\cal L}^{em}_M = -{1\over 16\pi}F^{\mu\nu}F_{\mu\nu},$$
$${\cal L}^{pf}_M= -A^{\mu}J_{\mu}  -
\sum_{i} {m_{i}\over \sqrt{|g|}e^{-3\Theta} }{ds_{i}\over dt}
\delta(\vec x - \vec x_{i}(t)),$$  and
$$J^{\mu}=\sum_{i}{q_{i}\delta(\vec x - \vec x_{i}(t))
\over \sqrt{|g|}e^{-3\Theta}}{dx^{\mu}\over dt} $$
\noindent(for Lagrangian density for particles in Saa's model
see \cite{Fiziev1}). It can be shown that
\ben
{\cal L}^{pf}_M - {\sfrac 1 3}{\delta{\cal L}^{pf}_M \over \delta \Theta}=0.
\een
Consequently we obtain
\ben
\Lambda = 3\int\left({\cal L}_M -
{\sfrac 1 3}{\delta{\cal L}_M \over \delta \Theta}\right)
\sqrt{|g|}e^{-3\Theta}d^3x =3\int\left({\cal L}^{em}_M -
{\sfrac 1 3}{\delta{\cal L}^{em}_M \over \delta \Theta}\right)
\sqrt{|g|}e^{-3\Theta}d^3x =          \nonumber
\een
\ben
-3\int {1 \over 8\pi}\left({\vec E}^2 - {\vec H}^2\right)
\sqrt{|g|}e^{-3\Theta}d^3x = -3\left({\cal E}_e - {\cal E}_m \right).
\een
Here ${\cal E}_e$ and ${\cal E}_m$ are correspondingly the electric and the
magnetic energy containt of the test body.
Therefore from (\ref{TBEM}) we have
\ben
{d^2X^{\beta}\over dt^2} =-3S^{\beta}(X){{\cal E}_e - {\cal E}_m \over M}.
\een
For aluminum and platinum the magnetic energy is small compared to the
electric one. The ratio ${{\cal E}_e \over M}$ for aluminum and platinum is
\cite{Ni}
\ben
\left({{\cal E}_e \over M}\right)_{Al} =1.7 * 10^{-3},
\een
\ben
\left({{\cal E}_e \over M}\right)_{Pt} =4.5 * 10^{-3}.
\een
On Earth we have
\ben
\left({d^2X_{i}\over dt^2}\right)_{Al} -
\left({d^2X_{i}\over dt^2}\right)_{Pt} =
-3\left(\left({{\cal E}_e \over M}\right)_{Al}  -
\left({{\cal E}_e \over M}\right)_{Pt} \right)S_{i}
=8.4 * 10^{-3} S_{i},
\een
$i$  runs from $1$ to $3$.

Out of the source the gravitational potential and the torsion vector are
related by  $S_{i}=K U{,i}$ (see (\ref{VS})) and therefore
\ben
\left({d^2X_{i}\over dt^2}\right)_{Al} -
\left({d^2X_{i}\over dt^2}\right)_{Pt} = 8.4 * 10^{-3} K U_{,i}.
\la{DA}
\een
According to the experiments of Roll-Krotkov-Dicke \cite{RKD} and
Braginsky-Panov \cite{BP} the acceleration of aluminum and platinum
don't differ by 1 part of $10^{11}$ or $10^{12}$  of $U_{,i}$ respectively
in the solar gravitational field. Taking into account that
$ {1 \over 3}\leq K \leq {1 \over 2}$  we conclude that  Saa's model
contradicts to these experiments.

It's not difficult to see that Saa's model contradicts to the experimental
data in E\"otvos experiment, too. Indeed, according to E\"otvos experiment the
acceleration of the bodies of different materials in Earth's gravitation field
don't differ by 1 part of $10^{11}$ of $g$ \cite{Will} (or $10^{12}$
according to \cite{SHAGHSS}) .
Using again relation (\ref{DA}) and inequality
$ {1 \over 3}\leq K \leq {1 \over 2} $
we see that Saa's model contradicts to E\"otvos experiment.

\bigskip
\bigskip

\section{Summary}

In this article we have examined the basic spherically symmetric
stationary state of stars in the Saa's model of gravity
with propagating torsion.

In the model under investigation there is no need to consider unknown charges
creating the torsion-dilaton field. Its source is the very spinless matter.
The whole geometry of the space-time (including metric and torsion)
is determined by the familiar properties of this matter.

The parameters of the vacuum solution are determined only by the spinless
matter without adopting an existence of new properties, too.
In contrast to the corresponding models in the general relativity
here we have two parameters $K$ and $a$ of the vacuum solutions.
The values of these parameters depend on the mass distribution in the star
which is related with the equation of state of the star's matter.
For a fixed equation of state both parameters become functions
only of the star mass, but these functions are not the same
for the different equations of state.
The first parameter $K$ being the ratio of the
magnitude of torsion-dilaton force and of the magnitude of gravitational
force for realistic equations of matter state takes values
in the interval $[{1\over 3},{1\over 2}]$, depending on the star's mass.
The second one -- $a$ is analogous to the gravitational radius in general
relativity and takes positive values depending on the value of the parameter
$K$ and on the value of the star's mass.

To be specific in the present article
we restrict our attention to the model of neutron stars where the effects
of nonlinearity are essential as in general relativity.
Numerical results and analytical considerations show that the space-time
torsion may have a significant role in their structure.
The new torsion force decrease in some extent the role of the gravity
in the star configuration and may lead to
an increasing or decreasing of the maximum neutron star mass depending on
the equation of state.

The complete investigation of the consistence of the whole Saa's model
of gravity with propagating torsion (including all type of physical fields)
with the reality is still an open problem.
The results of the present article may have not only independent value,
but are necessary for reaching the solution of this critical problem.
For example, after the first version of the present article was send
for publication a new results based on it which show the inconsistency
of Saa's model with solar system gravitational experiments were found
and published independently \cite{FY}.

Saa's model is a simple model based on pure geometrical reasons which
allows to overcome a basic difficulties of the old models of gravity with
torsion:
the inconsistency of the application of the minimal coupling principle in
action principle and directly in the equations of motion \cite{Saa1}.
Moreover, it leads to propagating torsion which is another important physical
property of this model. Unfortunately, it is not compatible with basic
physical experiments. This means that one has to modify this model preserving
its important new physical properties in a proper way to comply with real
physics.
A new investigations in this direction are in progress (see for example
\cite{F2}, \cite{F3}).

Nevertheless the model under consideration contradicts to basic
experiments its detailed investigation is quite instructive because
Saa's model is a special case of scalar-tensor theory of gravity with a
nonminimal coupling of the scalar torsion-dilaton field with matter.
A similar, but more general coupling of string-dilaton can be expected in
the string theory and it leads different physical effects.
For example, in the article by Damour and Polyakov \cite{DP} as a
consequences of nonminimal coupling between matter and dilaton was derived
a violation of the equivalence principle, as far as some string-dilaton
effects in the early universe \cite{DP}. Unfortunately, the present days string theory is not able to
predict definitely the form of the coupling between string-dilaton and the
real matter.
In contrast, Saa's model is the only one we know with complete determined
interaction of the (torsion) dilaton with all kinds of matter.
This interaction is similar to the one expected in other models.
In the present article it is shown at first that the nonminimal coupling of
the dilaton field with the matter will emerge in extreme conditions
in a neutron star and that it leads to clear new physical phenomena
of different kind, some details of which may depend on the equation of matter
state. Most probably similar effects will appear in other possible
modifications of the theory of torsion-dilaton, as far as in the other types
of theories of dilaton. For example the neutron stars may give us a way to
a real physics in string theories if the string-dilaton interactions with
real matter will be established.
Hence, the most important conclusion of present article is that looking for
physical manifestation of the dilaton one has to investigate in details its
influence on the neutron star structure.

\bigskip
\bigskip
\bigskip

\noindent{\Large\bf Acknowledgments}

\bigskip
\bigskip
\bigskip
We are deeply grateful to the unknown referees who suggested to consider
different types of masses of the neutron star in Saa's model and
stability analysis of the neutron star via catastrophe theory, and
to add to the present paper results about the consistence of this model
with physical reality, as far as for pointing out the references \cite{DT},
\cite{TLS}, \cite{CS}, \cite{TSL}, \cite{HRR}, \cite{SG}, \cite{Whinnett},
\cite{KMS1} and \cite{KMS2}.
The work on this article has been partially supported by
the Sofia University Foundation for Scientific Researches,
Contracts~No.No.~245/97,~257/97,
and by
the Bulgarian National Foundation for Scientific Researches,
Contract~No.~F610/97.
One of us (PF) is grateful to the leadership of the Bogoliubov Laboratory
of Theoretical Physics, JINR, Dubna, Russia for hospitality and
working conditions during his stay there in the summer of 1998
when a part of this investigation has been completed.

\end{document}